\documentclass[10pt,twocolumn,letterpaper]{article}

\usepackage{iccv}
\usepackage{times}
\usepackage{epsfig}
\usepackage{graphicx}
\usepackage{amsmath}
\usepackage{amssymb}
\usepackage{xurl}

\DeclareMathOperator*{\argmin}{arg\,min}

\usepackage{multirow}
\usepackage[table]{xcolor}
\usepackage{hhline}
\usepackage{booktabs}

\usepackage[pagebackref=true,breaklinks=true,letterpaper=true,colorlinks,bookmarks=false]{hyperref}

\iccvfinalcopy 

\def\iccvPaperID{3501} 

\ificcvfinal\fi

\begin{document}

\title{Learning to Reduce Defocus Blur by Realistically Modeling Dual-Pixel Data}

\author{Abdullah Abuolaim$^1$\footnotemark[1]
\qquad
Mauricio Delbracio$^2$
\qquad
Damien Kelly$^2$
\qquad
Michael S. Brown$^1$
\\
Peyman Milanfar$^2$\\
$^1$York University \qquad \qquad $^2$Google Research
}

\maketitle
\ificcvfinal\thispagestyle{empty}\fi

\begin{abstract}
Recent work has shown impressive results on data-driven defocus deblurring using the two-image views available on modern dual-pixel (DP) sensors.  One significant challenge in this line of research is access to DP data.  Despite many cameras having DP sensors, only a limited number provide access to the low-level DP sensor images. In addition, capturing training data for defocus deblurring involves a time-consuming and tedious setup requiring the camera's aperture to be adjusted. Some cameras with DP sensors (e.g., smartphones) do not have adjustable apertures, further limiting the ability to produce the necessary training data.  We address the data capture bottleneck by proposing a procedure to generate realistic DP data synthetically.  Our synthesis approach mimics the optical image formation found on DP sensors and can be applied to virtual scenes rendered with standard computer software.   Leveraging these realistic synthetic DP images, we introduce a recurrent convolutional network (RCN) architecture that improves deblurring results and is suitable for use with single-frame and multi-frame data (e.g., video) captured by DP sensors. Finally, we show that our synthetic DP data is useful for training DNN models targeting video deblurring applications where access to DP data remains challenging.
\end{abstract}


\begin{figure}[t]
\includegraphics[width=\linewidth]{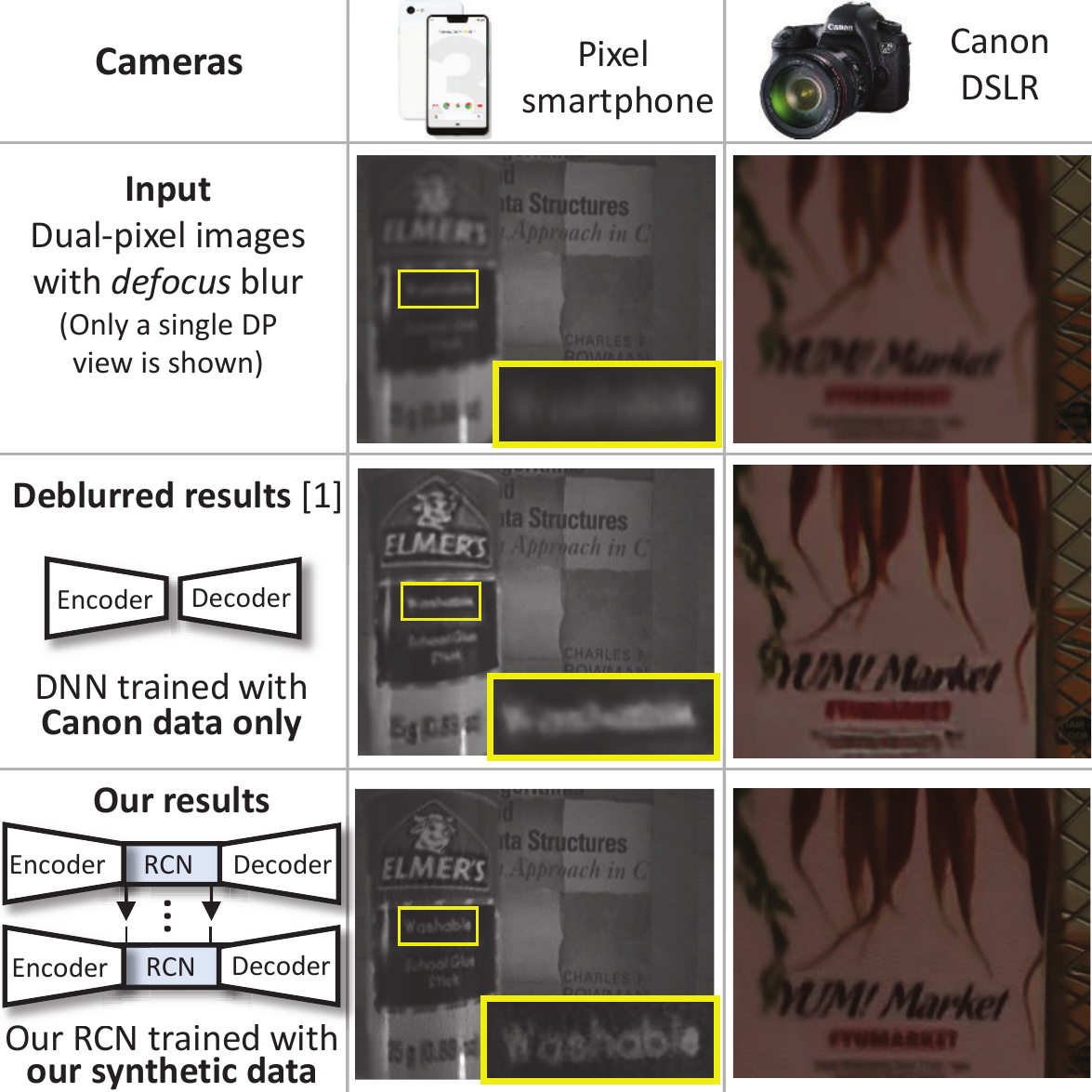}
\caption{Deblurring results on images from a Pixel 4 smartphone and a Canon 5D Mark IV. \textbf{Third row:} results of the DNN proposed in~\cite{abuolaim2020defocus} trained with DP data from the Canon camera. \textbf{Last row:} results from our proposed network trained on {\it synthetically generated data only}.}\label{fig:teaser}
\end{figure}

\section{Introduction and related work}
\footnotetext[1]{This work was done while Abdullah was an intern at Google.}
Defocus blur occurs in scene regions captured outside the camera's depth of field (DoF). Although the effect can be intentional (e.g., the bokeh effect in portrait photos), in many cases defocus blur is undesired as it impacts image quality due to the loss of sharpness of image detail (e.g., Fig.~\ref{fig:teaser}, second row). Recovering defocused image details is challenging due to the spatially varying nature of the defocus point spread function (PSF)~\cite{levin2011understanding,tang2012utilizing}, which not only is scene depth dependent, but also varies based on the camera aperture, focal length, focus distance, radial distortion, and optical aberrations. Most existing deblurring methods~\cite{d2016non,karaali2017edge,lee2019deep,park2017unified,shi2015just} approach the defocus deblurring problem by first estimating a defocus image map.  The defocus map is then used with an off-the-shelf non-blind deconvolution method (e.g., \cite{fish1995blind,krishnan2009fast}).  This strategy to defocus deblurring is greatly limited by the accuracy of the estimated defocus map.

\begin{figure*}[t]
\centering
\includegraphics[width=\linewidth]{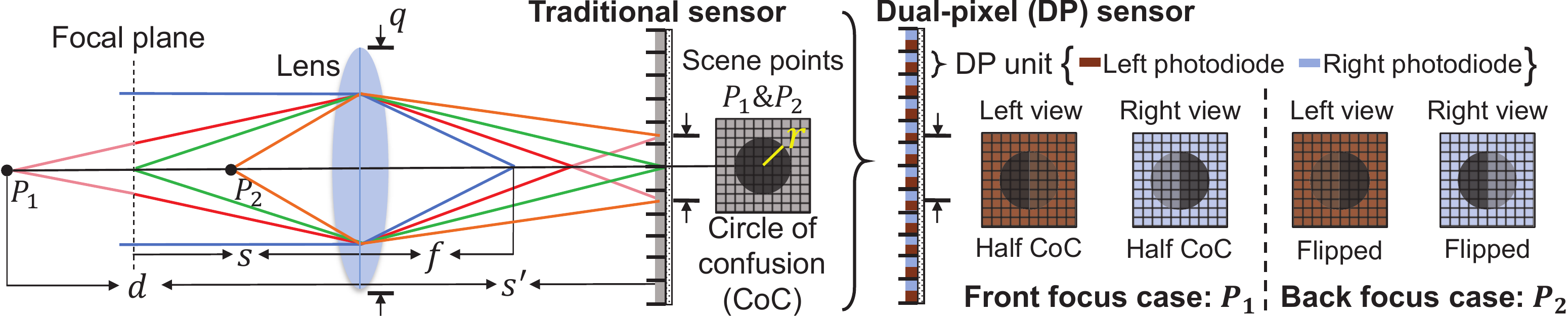}
\caption{Thin lens model illustration and dual-pixel image formation. The circle of confusion (CoC) size is calculated for a given scene point using its distance from the lens, camera focal length, and aperture size. On the two dual-pixel views, the half-CoC PSF flips if the scene point is in front or back of the focal plane.}\label{fig:thinlens}
\end{figure*}

Recently, work in~\cite{abuolaim2020defocus} was the first to propose an interesting approach to the defocus deblurring problem by leveraging information available on dual-pixel (DP) sensors found on most modern cameras.  The DP sensor was originally designed to facilitate auto-focusing~\cite{abuolaim2020online,abuolaim2018revisiting,jang2015sensor}; however, researchers have found DP sensors to be useful in broader applications, including depth map estimation~\cite{garg2019learning,pan2021dual,punnappurath2020modeling,fanellodu2net}, defocus deblurring~\cite{abuolaim2021ntire, vo2021attention,lee2021iterative,pan2021dual}, reflection removal~\cite{punnappurath2019reflection}, and synthetic DoF~\cite{wadhwa2018synthetic}.
DP sensors consist of two photodiodes at each pixel location effectively providing the functionality of a simple two-sample light-field camera (Fig.~\ref{fig:thinlens}, DP sensor). Light rays coming from scene points within the camera's DoF will have no difference in phase, whereas light rays from scene points outside the camera's DoF will have a relative shift that is directly correlated to the amount of defocus blur. Recognizing this, the work in~\cite{abuolaim2020defocus} proposed a deep neural network (DNN) framework to recover a deblurred image from a DP image pair using ground-truth data captured from a Canon DSLR.

\begin{figure}[t]
\centering
\includegraphics[width=\linewidth]{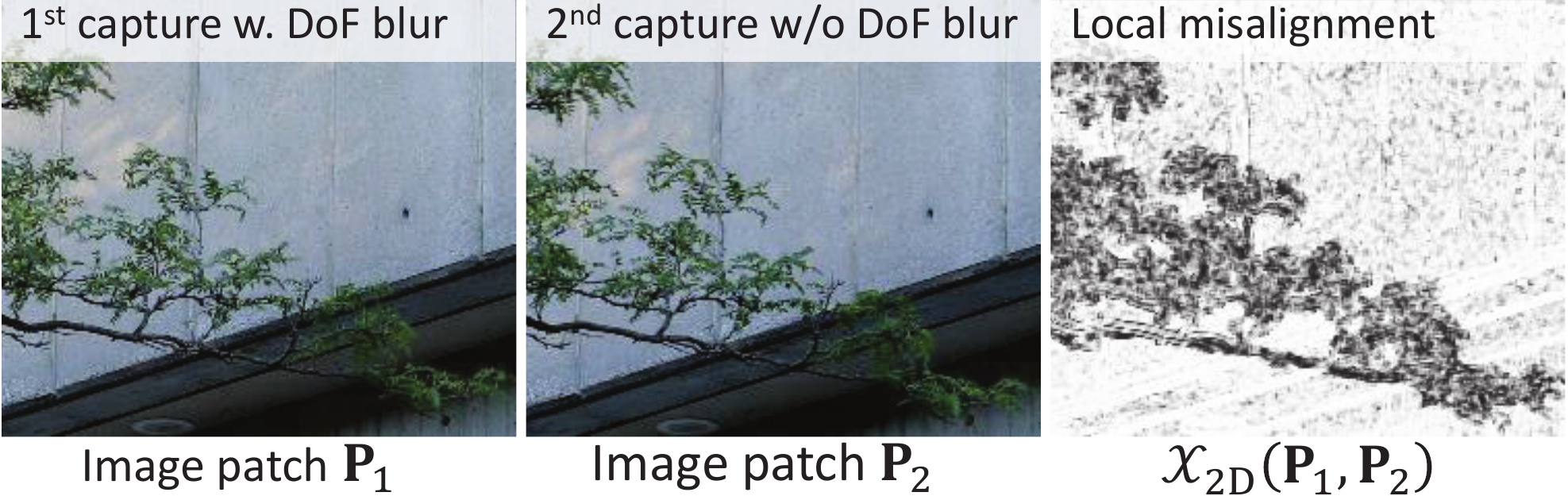}
\caption{Misalignment in the Canon DP dataset~\cite{abuolaim2020defocus} due to the physical capture procedure. 
Patches $\mathbf{P}_1$ and  $\mathbf{P}_2$ are cropped from in-focus regions from two captures: the first using a wide-aperture (w/ defocus blur) and the second capture using a narrow-aperture (w/o defocus blur).  The 2nd capture is intended to serve as the ground truth for this image pair. The third column shows the 2D cross correlation between the patches $\mathcal{X}_\mathrm{2D}(\mathbf{P}_1,\mathbf{P}_2)$, which reveals the local misalignment that occurs in such data capture.}\label{fig:misalignment}
\vspace{-2mm}
\end{figure}

Although the work of~\cite{abuolaim2020defocus} demonstrates state-of-the-art deblurring results, it is restricted by the requirement for accurate ground truth data, which requires DP images to be captured in succession at different apertures. As well as being labor intensive, the process requires careful control to minimize the exposure and motion differences between captures (e.g., see local misalignment in Fig.~\ref{fig:misalignment}).  Another significant drawback is that data capture is limited to a {\it single} commercial camera, the Canon 5D Mark IV, the only device that currently provides access to raw DP data and  has a controllable aperture.

While datasets exist for defocus estimation, including CUHK~\cite{shi2014discriminative}, DUT~\cite{zhao2018defocus}, and SYNDOF~\cite{lee2019deep}, as well as lightfield datasets for  defocus deblurring ~\cite{d2016non} and depth estimation \cite{hazirbas2018deep,srinivasan2017learning}, none provide DP image views. The work of \cite{abuolaim2020defocus} is currently the only source of ground truth DP data suitable for defocus deblur applications but is limited to a single device. This lack of data is a severe limitation to continued research on data-driven DP-based defocus deblurring, in particular to applications where the collection of ground truth data is not possible (e.g., fixed aperture smartphones).

\noindent{\textbf{Contributions.}}~This work aims to overcome the challenges in gathering ground-truth DP data for data-driven defocus deblurring. In particular, we propose a generalized model of the DP image acquisition process that allows realistic generation of synthetic DP data using standard computer graphics-generated imagery. We demonstrate that we can achieve state-of-the-art defocus deblurring results using synthetic data only (see Fig.~\ref{fig:teaser}), as well as complement real-image data sets through data augmentation. To demonstrate the generality of the model, we explore a new application domain of video defocus deblurring using DP data and propose a recurrent convolutional network (RCN) architecture that scales from single-image deblurring to video deblurring applications. Additionally, our proposed RCN addresses the issue of patch-wise training by incorporating radial distance learning and improves the deblurring results with a novel multi-scale edge loss. Our comprehensive experiments demonstrate the power of our synthetic DP data generation procedure and show that we can achieve the state-of-the-art results quantitatively and qualitatively with a novel network design trained with this data.

\begin{figure*}[t]
\centering
\includegraphics[width=\linewidth]{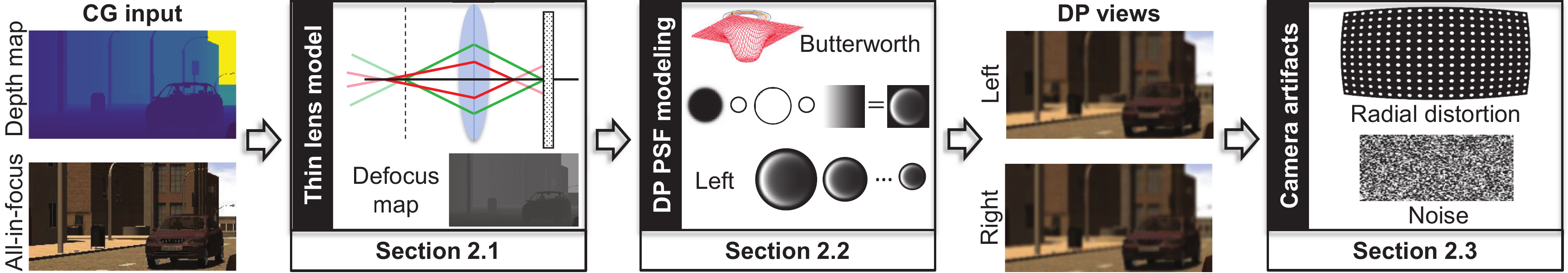}
\caption{An overview of our framework used to synthetically generate dual-pixel (DP) views. Our approach starts with computer-generated (CG) imagery produced with a standard computer graphics package.  Starting from this data, we model scene defocus, PSFs related to the DP sensor image formation, and additional artifacts, including radial distortion and sensor noise.}\label{fig:overview}
\end{figure*}

\section{Modeling defocus blur in dual-pixel sensors}
\label{sec:dpmodeling}

Synthetically generating realistic blur has been shown to improve data-driven approaches to both defocus map ~\cite{lee2019deep} and depth map estimation ~\cite{maximov2020focus}. We follow a similar approach in this work but tackle the problem of generating realistic defocus blur targeting DP image sensors. For this, we comprehensively model the complete DP image acquisition process with spatially varying PSFs, radial lens distortion, and image noise. Fig.~\ref{fig:overview} shows an overview of our DP data generator that enables the generation of realistic DP views from an all-in-focus image and corresponding depth map. 

\begin{figure*}[t]
\centering
\includegraphics[width=\linewidth]{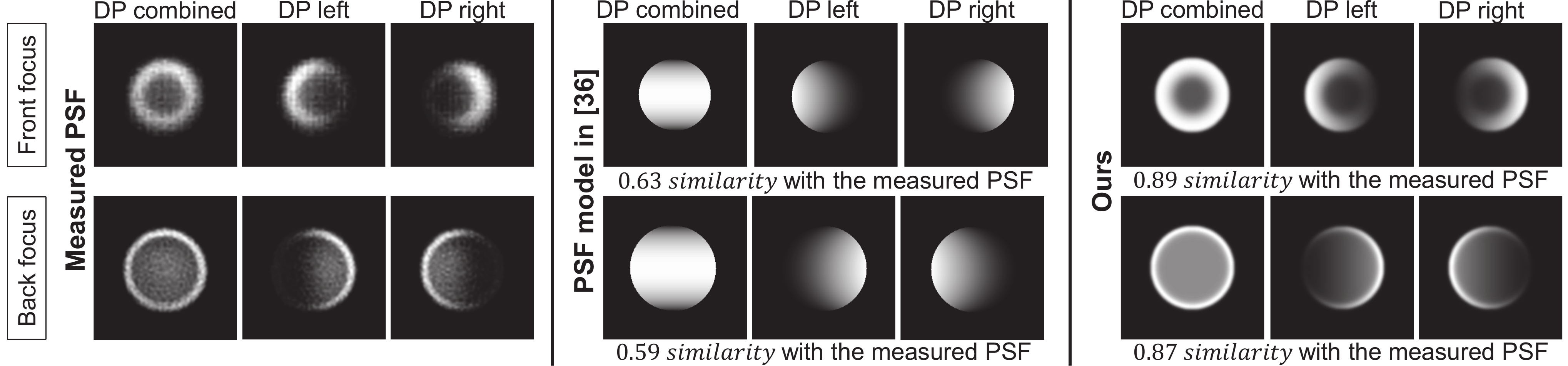}
\caption{Front and back focus DP PSFs. The similarity between two PSFs is measured by the 2D cross correlation. \textbf{Left}: Measured DP PSFs from the Canon 5D Mark IV DSLR. \textbf{Middle}: DP PSFs as modeled by~\cite{punnappurath2020modeling}. \textbf{Right}: Our newly proposed model for DP PSFs based on a modified 2D Butterworth filter. Our modeling achieves higher correlation with the real-world measured PSFs.}\label{fig:dpPsfs}
\end{figure*}

\subsection{Thin lens model}\label{sec:thinLens}
\label{sec:thinlens}
We model our virtual camera optics using a thin lens model that assumes negligible lens thickness, helping to simplify optical ray tracing calculations~\cite{potmesil1981lens}. Through this model, we can approximate the circle of confusion (CoC) that represents the PSF for a given point based on its distance from the lens and camera parameters (i.e., focal length, aperture size, and focus distance). This model is illustrated in Fig.~\ref{fig:thinlens}, where $f$ is the focal length, $s$ is the focus distance, and $F$ is the f-stop. The distance between the lens and sensor $s'$, and the aperture diameter $q$ are defined as $s'=\frac{f\;s}{s-f}$ and  $q=\frac{f}{F}$.
Then, the CoC radius $r$ of a scene point $P_1$ located at distance $d$ from the camera is:
\begin{equation}\label{eq:cocRadius}
r=\frac{q}{2} \times \frac{s'}{s} \times \frac{d-s}{d}.
\end{equation}

\subsection{Dual-pixel PSFs}

Recent work in~\cite{punnappurath2020modeling} introduced a model for approximating a PSF that occurs in the \emph{left} and \emph{right} views of a DP sensor using a nice symmetry property between the \emph{left} and \emph{right} PSFs. However, the model involved a single free parameter only that was directly correlated to the CoC size (Fig.~\ref{fig:dpPsfs}, middle column).  Though the model is able to capture the symmetry property observed in a real DP PSF, the overall PSF did not sufficiently reflect the true structure exhibited by real-world PSFs, as illustrated in Fig.~\ref{fig:dpPsfs}'s left column.  Real DP PSFs exhibit a donut-shaped depletion in the CoC that is attributed to optical aberrations~\cite{tang2012utilizing}.

To provide more realistic PSFs for the DP views, we introduce a parametric model based on the 2D Butterworth filter $\mathbf{B}$, defined as follows:
\begin{equation}\label{eq:butterworth}
\resizebox{.8\hsize}{!}{$\mathbf{B}(x,y) = \left(1+\left(\frac{D_o}{\sqrt{(x-x_o)^2+(y-y_o)^2}}\right)^{2n}\right)^{-1}$},
\end{equation}
where $n$ is the filter order, and $D_o$ is a parameter controlling the $3$dB cutoff position.
Aiming to capture the donut-shaped structure of the PSF, we define a parametric PSF model based on the Butterworth filter $\mathbf{B}$ as follows:
\begin{equation}\label{eq:ourPsf}
\mathbf{H}=\mathbf{B} \circ \mathbf{C}(x_o,y_o),
\end{equation}
where $\mathbf{C}$ represents a circular disk with radius $r$ matching the CoC radius as calculated in Eq.~\ref{eq:cocRadius}. The notation $\circ$ denotes the Hadamard product. Both $\mathbf{B}$ and $\mathbf{C}$ are centered at $(x_o,y_o)$. $D_o$ is a function of the radius $r$ and is controlled by the parameter $\alpha$. The values of $\mathbf{B}$ are re-scaled to $[\beta, 1]$, where the parameter $\beta > 0$ is introduced to control the minimum depletion at the kernel's center (which is always positive based on our observation of PSFs measured from real-world data).
With our proposed model, the parameterized PSF $\mathbf{H}$ has a sharp fall-off about the circumference. Therefore, we smooth $\mathbf{H}$ by convolving it with a Gaussian kernel of standard deviation $\kappa \times r$, where $0 < \kappa << 1$.

Our modeling of $\mathbf{H}$ represents the combined DP PSF, which is formed as $\mathbf{H}=\mathbf{H}_l+\mathbf{H}_r$,
where $\mathbf{H}_l$ and $\mathbf{H}_r$ are the \emph{left} and \emph{right} DP PSFs, respectively. Similar to the work in~\cite{punnappurath2020modeling}, we enforce the constraint of horizontal symmetry between $\mathbf{H}_l$ and $\mathbf{H}_r$, and express $\mathbf{H}_r$ as $\mathbf{H}_r=\mathbf{H}^f_l$, where $\mathbf{H}^f_l$ represents the \emph{left} PSF flipped about the vertical axis. $\mathbf{H}_l$ can be shown as $\mathbf{H}$ with a gradual fall-off towards the right direction (see front-focus DP left in Fig.~\ref{fig:dpPsfs}). Mathematically, we denote $\mathbf{H}_l$ as:
\begin{eqnarray}\label{eq:Hl}
&\mathbf{H}_l=\mathbf{H} \circ \mathbf{M}, \quad \mbox{s.t. } \ \mathbf{H}_l \geq \mathbf{0},\; \mbox{with} \sum \mathbf{H}_l = \frac{1}{2},
\end{eqnarray}
where $\mathbf{M}$ is a 2D ramp mask with a constant decay. This decay can be considered as an intensity fall-off (intensity/pixel) in a given direction. The direction is determined by the sign of the CoC radius calculated based on the thin lens model. The positive sign represents the front focus (i.e., blurring of objects behind the focal plane), whereas the negative sign represents the back focus (i.e., blurring of objects in front of the focal plane). Our PSF model, parameterized with five parameters, facilitates synthesizing PSF shapes more similar to what we measured in real cameras under different scenarios (see Fig.~\ref{fig:dpPsfs}, right column). From this model, we can generate a bank of representative PSFs based on actual observations from real cameras. Additional details about the calibration procedure, PSF estimation method, and parameter searching are provided in the supplemental material.

\subsection{Modeling additional camera artifacts}

\noindent{\textbf{Radial lens distortion.}}~Radial lens distortion occurs due to lens curvature imperfections causing straight lines in the real world to map to circular arcs in the image plane. This is a well-studied topic with many methods for modelling and correcting the radial distortion~(e.g.,~\cite{bukhari2013automatic,fitzgibbon2001simultaneous,hartley2007parameter,prescott1997line}). In our framework, we consider applying radial distortion to the synthetically generated images to mimic this effect found in real cameras. We adopt the widely used division model introduced in~\cite{fitzgibbon2001simultaneous}, as follows:
\begin{eqnarray}\label{eq:radialDistortionX}
(x_d,y_d)=(x_o,y_o) + \frac{(x_u-x_o,y_u-y_o)}{1+c_1 R^2+c_2 R^4+\cdots},
\end{eqnarray}
where $(x_u,y_u)$ and $(x_d,y_d)$ are the undistorted and distorted points respectively, and $c_i$ is the $i^\mathrm{th}$ radial distortion coefficient. $R$ is the radial distance from the image plane center $(x_o,y_o)$. This model enables different types of radial distortion, including barrel and pincushion. We generate representative radial distortion sets at different focal lengths found on cameras. A detailed description of this procedure is provided in the supplemental material.

\begin{figure*}[t]
\centering
\includegraphics[width=\linewidth]{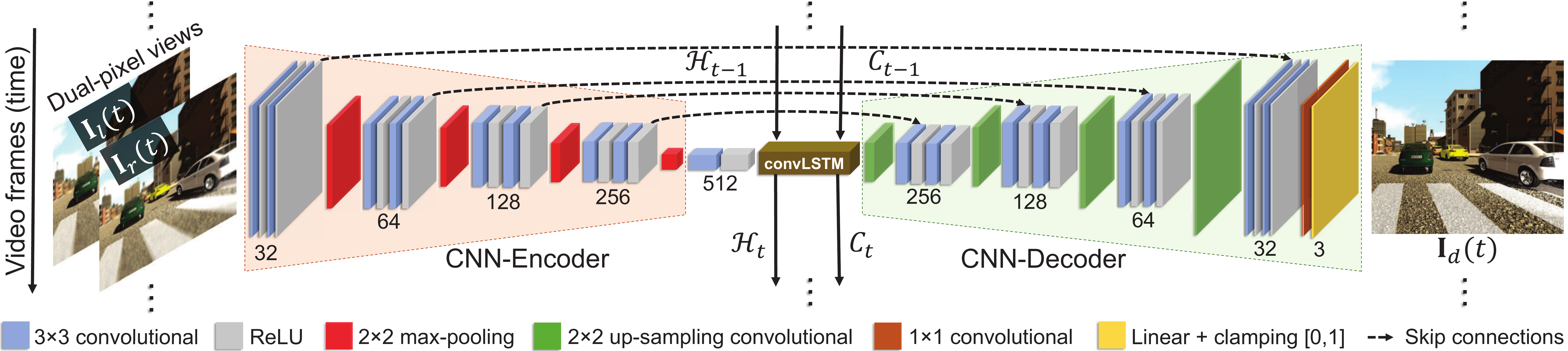}
\caption{Our recurrent dual-pixel deblurring (RDPD) architecture. Our model takes a blurred image sequence, where each image at time $t$ is fed as \emph{left} $\mathbf{I}_l(t)$ and \emph{right} $\mathbf{I}_r(t)$ DP views. The DP views are encoded at the encoder part to feed the convLSTM that outputs the hidden state $\mathcal{H}_t$ and memory cell $\mathcal{C}_t$ to the next time point. The convLSTM unit also outputs a feature map $o_t$ that is processed through the decoder part to give the deblurred sharp image $\mathbf{I}_d(t)$. Note: the number of output filters is shown under each convolution operation.}\label{fig:convLSTM}
\end{figure*}

\noindent{\textbf{Noise.}}~Image noise is the undesirable occurrence of random variations in intensity or color information in images. Our initial input is CG-generated data that is noise-free.  In order to synthesize realistic images, we add signal-dependent noise as the last step. We model the noise using a signal-dependent Gaussian distribution where the variance of noise is proportional to image intensity~\cite{foi2008practical,liu2007automatic}. Let $\mathbf{I}$ be the noiseless image and $\mathbf{N}$ a zero-mean Guassian noise layer; then our modeling of the signal-dependent Gaussian noise is $\mathbf{I}_\text{noise} = \mathbf{I} + \mathbf{I} \circ \mathbf{N}$,
where $\mathbf{N} \sim \mathcal{N}(\mathbf{0}, \sigma^2 \text{Id})$, and $\sigma$ controls the noise strength.

\section{Generating dual-pixel views}

In this section we introduce the synthetic dataset used, followed by a description of the procedure to synthetically generate the DP \emph{left} and \emph{right} views.  The source of our synthetic example data comes from the street view SYNTHIA dataset~\cite{HernandezBMVC17}, which contains image sequences of photo-realistic GC-rendered  images from a virtual city. Each sequence has 400 frames on average. The dataset contains a large diversity in scene setups, involving many objects, cities, seasons, weather conditions, day/night time, and so forth. The SYNTHIA dataset also includes the depth-buffer and labelled segmentation maps. In our framework, we use the depth map to apply synthetic defocus blur in the process of generating the DP views.

To blur an image based on the computed CoC radius $r$, we first decompose the image into discrete layers according to per-pixel depth values, where the maximum number of layers is set to $500$. Then, we convolve each layer with our parameterized PSF, blurring both the image and mask of the depth layer. Next, we alpha-blend the blurred layer images in order of back-to-front, using the blurred masks as alpha values. For each all-in-focus video frame $\mathbf{I}_s$, we generate two images -- namely, the \emph{left} $\mathbf{I}_l$ and \emph{right} $\mathbf{I}_r$ sub-aperture DP views -- as follows (for simplicity, let $\mathbf{I}_s$ be a patch with all pixels from the same depth layer):
\begin{equation} \label{eqn:dpViews}
\mathbf{I}_l = \mathbf{I}_s \ast \mathbf{H}_l, \qquad
\mathbf{I}_r = \mathbf{I}_s \ast \mathbf{H}_r,
\end{equation}
where $\ast$ denotes the convolution operation. Afterwards, the radial distortion is applied on $\mathbf{I}_l$, $\mathbf{I}_r$, and $\mathbf{I}_s$ based on the camera's focal length. Finally, we add signal-dependent noise layers (i.e., $\mathbf{N}_l$ and $\mathbf{N}_r$) for the two DP views that have the same $\sigma$, but are drawn independently. The final output defocus blurred image $\mathbf{I}_b$ is equal to $\mathbf{I}_l + \mathbf{I}_r$.

Our synthetically generated DP views exhibit a similar focus disparity to what we find in real data, where the in-focus regions show no disparity and the out-of-focus regions have defocus disparity. In the supplemental material, we provide animated examples that alternate between the \emph{left} and \emph{right} DP views to assist in visual comparisons between synthetic and real data.

\section{Defocus deblurring image sequences}

With the ability to generate synthetic DP data, we can shift our attention to training new RCN-based architectures addressing image sequences (e.g., video) captured with DP sensors. This is possible only by using our synthetic DP data, as no current device allows video DP data capture.  As we will show, our method can be used for both image sequences and single-image inputs. In the context of image sequences, the amount of defocus blur changes based on the motion of the camera and scene's objects over time. In the presence of such motion, sample depth variation over a sequence of frames provides useful information for deblurring. Our work is the first to explore the domain of defocus deblurring on image sequences (e.g., video). 

We adopt a data-driven approach for correcting defocus blur. We leverage a symmetric encoder-decoder CNN-based architecture with skip connections between corresponding feature maps~\cite{mao2016image,ronneberger2015u}. Skip connections are widely used in encoder-decoder CNNs and have been found to be effective for image deblurring tasks~\cite{abuolaim2020defocus,gast2019deep}. Our proposed network is also coupled with convLSTM units~\cite{tao2018scale,wang2018revisiting,xingjian2015convolutional} to better learn temporal dependencies between multiple frames and to allow variable sequence size. With convLSTM units, the same network remains fully convolutional and can successfully deblur a single image or a sequence of images of arbitrary number. Fig.~\ref{fig:convLSTM} shows a detailed overview of our proposed CNN-convLSTM architecture, which we refer to as {\it recurrent dual-pixel deblurring} (RDPD).

Our architecture is similar to the one in~\cite{abuolaim2020defocus}, but with the following modifications: (1) convLSTM units are added to the network bottleneck, (2) we train the network using the radial distance patch to address the patch-wise training issue, (3) we introduce a multi-scale edge loss function that helps in recovering sharp edges, (4) the number of nodes are reduced to half at each block to make the model lighter, and (5) the last layer is replaced by a linear layer with a [0,1] clamping as it is found to be more effective in~\cite{gast2019deep}.

\begin{table*}[t]
\small 
\centering
\caption{Results on the Canon DP dataset from~\cite{abuolaim2020defocus}. DPDNet is the pre-trained model on Canon data provided by~\cite{abuolaim2020defocus}. DPDNet+ and our RDPD+ are trained with Canon and our synthetically generated DP data. Bold numbers are the best and highlighted in green. The second-best performing results are highlighted in yellow. The testing set consists of 37 indoor and 39 outdoor scenes.}
\scalebox{0.96}{
\begin{tabular}{l|c|c|c|c|c|c|c|c|c|c|c}
\toprule
\multicolumn{1}{c|}{} & \multicolumn{3}{c|}{\textbf{Indoor}} & \multicolumn{3}{c|}{\textbf{Outdoor}} & \multicolumn{4}{c|}{\textbf{Indoor $\&$ Outdoor}} & \\ \cline{2-11}
\multicolumn{1}{c|}{\multirow{-2}{*}{\textbf{Method}}} & PSNR $\uparrow$ & SSIM $\uparrow$ & MAE $\downarrow$ & PSNR $\uparrow$ & SSIM $\uparrow$ & MAE $\downarrow$ & PSNR $\uparrow$ & SSIM $\uparrow$ & MAE $\downarrow$ & NIQE $\downarrow$ & \multicolumn{1}{|c}{\multirow{-2}{*}{\textbf{Time $\downarrow$}}} \\ \toprule

EBDB~\cite{karaali2017edge}	& 25.77 & 0.772 &0.040 & 21.25 & 0.599 & 0.058 & 23.45 & 0.683 & 0.049 & 5.42 & 929.7\\ \hline

DMENet~\cite{lee2019deep}	& 25.70 & 0.789 & 0.036 & 21.51 & 0.655 & 0.061 & 23.55 & 0.720 & 0.049 & 4.85 & 613.7\\ \hline

JNB~\cite{shi2015just}		& 26.73 & 0.828 & 0.031 & 21.10 & 0.608 & 0.064 & 23.84 & 0.715 & 0.048 & 5.11 & 843.1\\ \midrule \midrule

DPDNet~\cite{abuolaim2020defocus}	& 27.48 & \cellcolor{yellow!35} 0.849 & 0.029 & \cellcolor{green!25}{\bf 22.90} & \cellcolor{green!25}{\bf 0.726} & \cellcolor{green!25}{\bf 0.052} & \cellcolor{yellow!35} 25.13 & \cellcolor{green!25}{\bf 0.786} & \cellcolor{yellow!35} 0.041 & 3.77 & 0.5\\ \hline

DPDNet+~\cite{abuolaim2020defocus}	& \cellcolor{yellow!35}27.65 & \cellcolor{green!25}{\bf 0.852} & \cellcolor{yellow!35}0.028 & 22.72 & \cellcolor{yellow!35}{0.719} & 0.054 & 25.12 & \cellcolor{yellow!35}{0.784} & 0.042 & \cellcolor{yellow!35}{3.73} & 0.5\\ \hline

Our RDPD+	& \cellcolor{green!25}{\bf 28.10} & 0.843 & \cellcolor{green!25}{\bf 0.027} & \cellcolor{yellow!35} 22.82 & 0.704 & \cellcolor{yellow!35} 0.053 & \cellcolor{green!25}{\bf 25.39} & 0.772 & \cellcolor{green!25}{\bf 0.040} & \cellcolor{green!25}{\bf 3.19} & \cellcolor{green!25}{\bf 0.3}\\ \hline
\end{tabular}
}
\label{tab:quantitaiveResultsCanon}
\end{table*}

\noindent{\textbf{RDPD architecture.}}~Given an input video of $j$ consecutive frames that have defocus blur $\{\{\mathbf{I}_l(t),\mathbf{I}_r(t)\},\cdots,\{\mathbf{I}_l(t+j),\mathbf{I}_r(t+j)\}\}$ (such that $\mathbf{I}_l(t)$ and $\mathbf{I}_r(t)$ are the DP views of the given frame at time $t$), we first obtain a sequence of compact convolutional features $\{X(t),\cdots,X(t+j)\}$ encoded at the CNN bottleneck -- namely, $X(t)=\textrm{CNN-Encoder}(\mathbf{I}_l(t),\mathbf{I}_r(t))$. Then, the features are fed to a convLSTM as shown in Fig.~\ref{fig:convLSTM}. We utilize the convLSTM to learn of the temporal dynamics of the sequential inputs. This is achieved by incorporating memory units with the gated operations. The convLSTM also preserves spatial information by replacing dot products with convolutional operations, which is essential for making spatially variant estimation  align with the spatially varying DP PSFs. We choose LSTM over RNN because standard RNNs are known to have difficulty in learning long-time dependencies~\cite{hochreiter2001gradient}, whereas LSTMs have shown the capability to learn long- and short-time dependencies~\cite{hochreiter1997long}.

For the input feature $X(t)$ at time $t$, our convLSTM leverages three convolution gates -- input $i_t$, output $o_t$, and forget $\mathcal{F}_t$ -- in order to control the signal flow within the cell. The convLSTM outputs a hidden state $\mathcal{H}_t$ and maintains a memory cell $\mathcal{C}_t$ for controlling the state update and output:
\begin{equation}\label{eqn:convLSTM1}
i_t = \Sigma(W^X_i\ast X_t + W^{\mathcal{H}}_i\ast \mathcal{H}_{t-1} + W^{\mathcal{C}}_i\circ \mathcal{C}_{t-1} + b_i),
\end{equation}
\begin{equation}\label{eqn:convLSTM2}
\mathcal{F}_t = \Sigma(W^X_{\mathcal{F}}\ast X_t + W^{\mathcal{H}}_{\mathcal{F}}\ast \mathcal{H}_{t-1} + W^{\mathcal{C}}_{\mathcal{F}}\circ \mathcal{C}_{t-1} + b_{\mathcal{F}}),
\end{equation}
\begin{equation}\label{eqn:convLSTM3}
o_t = \Sigma(W^X_o\ast X_t + W^{\mathcal{H}}_o\ast \mathcal{H}_{t-1} + W^{\mathcal{C}}_o\circ \mathcal{C}_{t-1} + b_o),
\end{equation}
\begin{equation}\label{eqn:convLSTM4}
\mathcal{C}_t = \mathcal{F}_t \circ \mathcal{C}_{t-1} + i_t \circ\;\tau(W^X_{\mathcal{C}}\ast X_t + W^{\mathcal{H}}_{\mathcal{C}}\ast \mathcal{H}_{t-1} +b_{\mathcal{C}}),
\end{equation}
\begin{equation}\label{eqn:convLSTM5}
\mathcal{H}_t=o_t \circ  \tau(\mathcal{C}_t),
\end{equation}
where the $W$ terms denote the different weight matrices, and the $b$ terms represent the different bias vectors. $\Sigma$ and $\tau$ are the activation functions of logistic sigmoid and hyperbolic tangent, respectively. Afterwords, the output deblurred image $\mathbf{I}_d$ is obtained by decoding $o_t$ through the decoder part of our encoder-decoder CNN as follows:
\begin{equation}\label{eqn:convLSTM6}
\mathbf{I}_d(t)=\textrm{CNN-Decoder}(o_t).
\end{equation}

\noindent{\textbf{Radial distance patch.}}~Radial distortion and lens aberration make the PSFs vary in radial directions away from the image center. Similar to~\cite{abuolaim2020defocus,park2017unified}, we perform patch-wise training to avoid the redundancies of full image training and ensure that the input has enough variance. However, this approach breaks the spatial correlation between the image patches as they are fed independently with no knowledge of their location on the image plane.  As a result, in addition to the six-channel RGB DP views, we include a single-channel patch that represents the relative radial distance.

\noindent{\textbf{Multi-scale edge loss.}}~In addition to the MSE loss, we introduce a multi-scale edge loss using a Sobel gradient to guide the network to encourage sharper edges. Our new loss is similar in principle to the single-scale (i.e., $3\times3$) Sobel loss used in~\cite{lu2019single}, but we modified this loss in two ways: first, we added multiple scales of the Sobel operator (i.e., kernel sizes) in order to capture different edge sizes. Second, we minimized for the horizontal and vertical directions separately, to concentrate more on the direction that is perpendicular to the imaging sensor orientation. For our multi-scale modified edge loss, the vertical $G^x$ and horizontal $G^y$ derivative approximations of the deblurred output $\mathrm{I}_d$ and its ground truth $\mathrm{I}_s$ are:
\begin{eqnarray}\label{eqn:gradients}
G^x_d=\mathrm{I}_d \ast S^x_{m\times m}, \qquad G^y_d=\mathrm{I}_d \ast S^y_{m\times m},\\
G^x_s=\mathrm{I}_s \ast S^x_{m\times m}, \qquad G^y_s=\mathrm{I}_s \ast S^y_{m\times m},
\end{eqnarray}
where $S^x_{m\times m}$ and $S^y_{m\times m}$ are the vertical and horizontal Sobel operators of size $m$, respectively. The derivative operations are performed at multiple filter sizes. Our new edge loss $\mathcal{L}_{\text{edge}}$ is the mean of multiple scales for each direction $x/y$ and denoted as:
\begin{eqnarray}\label{eqn:sobelLoss}
\mathcal{L}^{\{x,y\}}_{\text{edge}}=\mathbb{E}[\textrm{MSE}(G^{\{x,y\}}_s,G^{\{x,y\}}_d)].
\end{eqnarray}
Then the final loss function $\mathcal{L}$ is:
\begin{equation}\label{eqn:allLoss}
\mathcal{L}=\mathcal{L}_{\textrm{MSE}}+\lambda_x\mathcal{L}^x_{\text{edge}}+\lambda_y\mathcal{L}^y_{\text{edge}},
\end{equation}
such that $\mathcal{L}_{\textrm{MSE}}$ is the typical MSE loss between the output estimated $\mathrm{I}_d$ and its ground truth $\mathrm{I}_s$. The $\lambda$ terms are added to control our final loss.

\begin{figure*}[t]
\centering
\includegraphics[width=\linewidth]{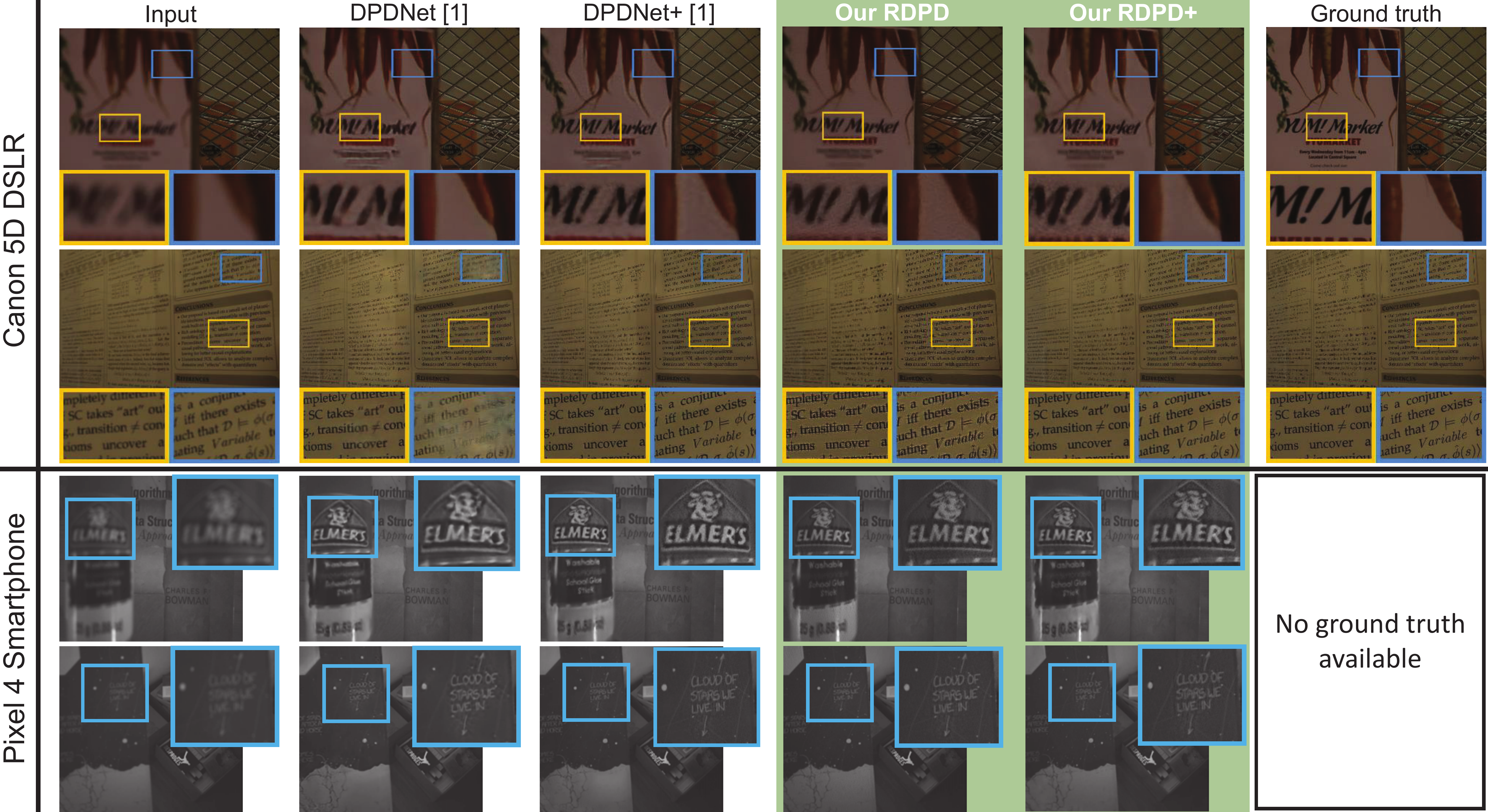}
\caption{Qualitative results. DPDNet~\cite{abuolaim2020defocus} is trained on Canon DP data. RDPD is our method trained on synthetically generated DP data only. DPDNet+ and RDPD+ are trained on {\it both} Canon and synthetic DP data. In general, RDPD and RDPD+ are able to recover more image details. Interestingly, RDPD trained on synthetic data generalizes well to real data from the two tested cameras. Note that there is no ground truth sharp image for Pixel 4, due to the fact smartphones have fixed aperture and thus a narrow-aperture image cannot be captured to serve as a ground truth image. Additionally, we note that the DP data currently available from the Pixel smartphones are not full-frame, but are limited to only one of the green channels in the raw-Bayer frame.
} \label{fig:qualResBoth}
\end{figure*}
\section{Experiments}
We evaluate our proposed RDPD and other existing defocus deblurring methods: the DP deblurring network (DPDNet)~\cite{abuolaim2020defocus}, the edge-based defocus blur (EBDB)~\cite{karaali2017edge}, the defocus map estimation network (DMENet)~\cite{lee2019deep}, and the just noticeable blur (JNB)~\cite{shi2015just} estimation. DPDNet~\cite{abuolaim2020defocus} is the only method that utilizes DP data for deblurring, and the others~\cite{karaali2017edge,lee2019deep,shi2015just} use only a single image as input (i.e., $\mathrm{I}_l$) and estimate the defocus map in order to feed it to an off-the-shelf deconvolution method (i.e.,~\cite{fish1995blind,krishnan2009fast}). EBDB~\cite{karaali2017edge} and JNB~\cite{shi2015just} are not learning-based methods; thus, we can test them directly. For the learning-based DMENet method, we cannot retrain it with the Canon data~\cite{abuolaim2020defocus}, as it does not provide the ground truth defocus map. However, with our data generator we are able to generate defocus maps, which allows us to retrain DMENet with our synthetically generated data.

\noindent{\textbf{Settings to generate DP data.}}~For our DP data generator, we define five camera parameter sets -- namely, $\{4,5,6\},\{5,8,6\},\{7,5,8\},\{10,13,12\},\{22,10,30\}$ -- such that each set represents focal length, aperture size, and focus distance. Given the depth range found in the SYNTHIA dataset~\cite{HernandezBMVC17}, these camera sets cover a wide range of front- as well as back-focus CoC sizes. For each image sequence in the SYNTHIA dataset, we generate five sequences based on the predefined camera sets. The radial distortion coefficients are set accordingly for each camera set. For the DP PSFs, we generate many representative PSF shapes by varying the parameters in the given ranges $n\in\{3,6,9\}$,  $\alpha\in\{0.4,0.6,0.8,1\}$, $\beta\in\{0.1,0.2,0.3,0.4\}$, and $\kappa=0.14$. Image noise layer strength is chosen randomly, where $\sigma \in \{5e^{-2},5.5e^{-2},\cdots, 5e^{-1}\}$. These parameters are set empirically to model real camera hardware. More detail is provided in the supplemental material.

We divide the SYNTHIA dataset~\cite{HernandezBMVC17} into training and testing sequences. We generate five sets of blurred images for each image sequence. In total, we synthesize 2023 training and 201 testing blurred DP views. Though our synthetic DP data generator enables an unlimited number of images to be generated, we found this number of images sufficient for training. In addition to our synthetically generated DP data, we use the DP ground truth data from~\cite{abuolaim2020defocus} with $300$ training, $74$ validation, and $76$ testing pairs of blurred images (with DP views) and corresponding sharp images.

\noindent{\textbf{RDPD settings and training procedure.}}~We set the size of the convLSTM to $512$ units. For patch-wise training, we fix the size of input and output layers to $512\times512\times7$ and $512\times512\times3$, respectively. We initialize the weights of the convolutional layers using He’s initialization~\cite{he2015delving} and use the Adam optimizer~\cite{kingma2014adam} to train the model. The initial learning rate is $5\times10^{-5}$, which is decreased by half every 40 epochs.

For domain generalization from synthetic to real data~\cite{guo2018learning,shrivastava2017learning}, we train our model iteratively using mini-batches of real (i.e., single image) as well as synthetic data (i.e., image sequence), where the patches are randomly cropped at each iteration. This type of iterative image/image sequence training becomes feasible since our recurrent model RDPD allows training and testing with any number of frames, and it does not need to be preset beforehand. We set the mini-batch size for the real data iteration to eight batches, because the dataset of real data has only single-image examples (i.e., no image sequences). For the synthetic data iteration, we set the mini-batch size to two sequences each of size four frames. We define three scales for our edge loss -- namely, $m \in \{3,7,11\}$. The $\lambda$ terms are found to be effective at $\lambda_x=0.03$ and $\lambda_y=0.02$.

To avoid overfitting, the dropout layer in the convLSTM is set to $0.4$. Our model converges after 140 epochs. Although we train on image patches, our RDPD (with convLSTM) is fully convolutional and enables testing on full-resolution inputs. To demonstrate the effectiveness of each component in our model, an ablation study of different training settings is provided in the supplemental material.

\noindent{\textbf{Single image results.}}~We evaluate our proposed RDPD against existing defocus deblurring methods for single-image inputs. For methods that utilize DP views for the input image (i.e., RDPD and DPDNet~\cite{abuolaim2020defocus}), we introduce variations on the training data used for more comprehensive evaluations. The variations are RDPD+ and DPDNet+ that are trained on both Canon DP data from~\cite{abuolaim2020defocus} combined with synthetic DP data generated by the process described in Sec.~\ref{sec:dpmodeling}. The RDPD without the $+$ sign is our baseline trained with synthetically generated DP data only. The DPDNet without the $+$ is trained on Canon data only.

\begin{figure}[t]
\centering
\includegraphics[width=\linewidth]{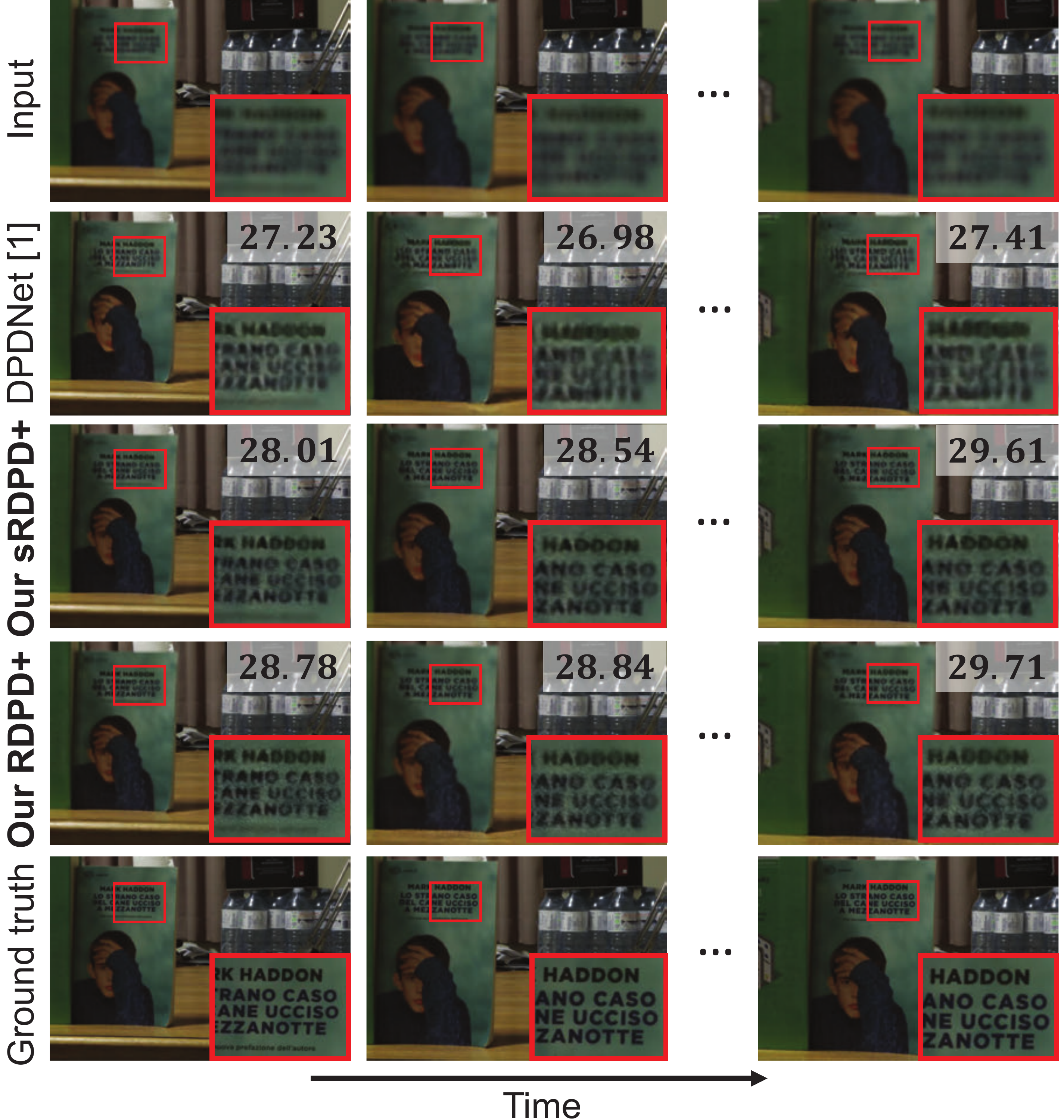}
\caption{Results on a Canon 5D DSLR image sequence. PSNR is shown for each deblurred image. sRDPD+ has a $0.4$dB lower PSNR on average when it is trained with a single frame compared to our multi-frame method -- that is, RDPD+.} \label{fig:canonSequence}
\vspace{-3mm}
\end{figure}

In Table~\ref{tab:quantitaiveResultsCanon}, we report quantitative results on real Canon DP data from~\cite{abuolaim2020defocus} using standard metrics -- namely, MAE, PSNR, SSIM, and time. We also report the Naturalness Image Quality (NIQE) metric of the output deblurred images with respect to a reference model derived from the DP GT images. In general, our RDPD+ has the best overall PSNR compared to other methods. Particularly, RDPD+ achieves the best PSNR and MAE for both indoor and combined categories, and all with our lighter-weight network that enabled the fastest inference time.

For the Outdoor dataset, the PSNR of RDPD+ is slightly lower (i.e., $0.08$dB) due to the fact that the Outdoor dataset is imperfect as a result of the capturing process (see  Fig.~\ref{fig:misalignment}). DP cameras do not enable simultaneous capture of the DP images and corresponding ground truth sharp image (i.e., the image pairs can be captured only in succession at different times). As a result, the Outdoor ground-truth is imperfect with small local motion and illumination variations. The Indoor ground-truth is captured in more controlled conditions and has fewer imperfections. The slightly better performance of DPDNet for outdoor scenes is because DPDNet is learning to compensate for imperfections in the Outdoor dataset. A key strength of our work is the ability to synthetically generate DP data that is not impeded by imperfections of manual capture. RDPD+ is trained on the Outdoor dataset as well as the synthetically generated data (without such imperfections) debiasing the result from the imperfect ground-truth. The consequence is reduced fidelity to the imperfect ground-truth (in terms of PSNR/SSIM), but better defocus deblur performance overall. Similar behavior is observed when DPDNet is trained with Canon data and our synthetically generated data (i.e., DPDNet+).

In Fig.~\ref{fig:qualResBoth}, we also provide qualitative results of RDPD compared to other methods on data captured by Canon DSLR and Pixel 4 cameras. In general, RDPD+ is able to recover more details from the input deblurred image. Additionally, Fig.~\ref{fig:qualResBoth} demonstrates that the baseline RDPD achieves good deblurring results on Canon and Pixel 4 data despite being trained with synthetic data only. This result demonstrates the accuracy of the proposed framework for synthetic DP data generation and the ability of the recurrent model to generalize to different cameras. It can also be seen that DPDNet+ has improved results compared to DPDNet, demonstrating the benefit gained by DPDNet+ through the addition of synthetic DP data on training. The supplemental material contains more quantitative results, visual comparisons, and animated deblurring examples for both Canon and Pixel 4 cameras.

\begin{table}[t]
\centering
\caption{Results on our synthetically generated DP data. sRDPD+ is a variation that is trained with single-frame data (\textbf{green}=best, \textbf{yellow}=second best). Our RDPD+, trained with image sequences, achieves the best results.}
\small
\scalebox{0.9}{
\begin{tabular}{l|c|c|c}
\toprule
\multicolumn{1}{c|}{\textbf{Method}} & PSNR $\uparrow$ & SSIM $\uparrow$ & MAE $\downarrow$\\
\toprule

DPDNet~\cite{abuolaim2020defocus}	& 26.38 &  0.782 & 0.034\\ \hline

DPDNet+~\cite{abuolaim2020defocus}	& 29.84 &  0.828 & 0.025\\ \hline

sRDPD+	& \cellcolor{yellow!35} 30.26 & \cellcolor{yellow!35} 0.849  & \cellcolor{yellow!35} 0.020 \\ \hline

RDPD+	& \cellcolor{green!25} {\bf 31.09} & \cellcolor{green!25} {\bf 0.861}  & \cellcolor{green!25} {\bf0.016}\\ \hline
\end{tabular}
}
\label{tab:quantitaiveResultsSynth}
\end{table}

\noindent{\textbf{Image sequence results.}}~Our RDPD is designed to handle input image sequences. Here, we investigate the improvement gained by training with image sequences vs. single frames. For this comparison, we introduce the RDPD+ variant sRDPD+, which is trained with single-frame inputs.

As previously mentioned, there is no camera that enables access to DP views for video data. Nevertheless, we mimic the same capturing procedure in~\cite{abuolaim2020defocus} in order to capture a sequence of images. We performed four captures of the same scene with small camera motion introduced between the captures. Each image has its own DP views and is captured at narrow and wide apertures. Fig.~\ref{fig:canonSequence} presents the results on the sequence of images. The effectiveness of training with image sequences with RDPD+ can be seen from the average PSNR gain (i.e., $+0.4$dB) compared to sRDPD+ trained using single-image inputs.

Table~\ref{tab:quantitaiveResultsSynth} shows the quantitative results on our synthetically generated DP image sequences. Our method RDPD+ (trained on multiple frames) achieves the best results as it utilizes the convLSTM architecture to better model the temporal dependencies in an image sequence. Recall that our RDPD network is lighter and has a much lower number of weights compared to DPDNet.
\section{Conclusion}
We proposed a novel framework to generate realistic DP data by modeling the image formation steps present on cameras with DP sensors.  Our framework helps to address the current challenges in capturing DP data.   Utilizing our synthetic DP data, we also proposed a new recurrent convolutional architecture that is designed to reduce defocus blur in image sequences. We performed a comprehensive evaluation of existing deblurring methods, and demonstrated that our synthetically generated DP data and recurrent convolutional model achieve state-of-the-art results quantitatively and qualitatively. Furthermore, our proposed framework demonstrates the ability to generalize across different cameras by training on synthetic data only. We believe our DP data generator will help spur additional ideas about defocus deblurring and applications that leverage DP data. Our dataset, code, and trained models are available at \url{https://github.com/Abdullah-Abuolaim/recurrent-defocus-deblurring-synth-dual-pixel}.
\\
\noindent{\textbf{Acknowledgments.}}
The authors would like to thank Shumian Xin, Yinxiao Li, Neal Wadhwa, and Rahul Garg for fruitful comments and discussions.
\clearpage

\setcounter{section}{0}
\setcounter{footnote}{0}
\setcounter{figure}{0}
\setcounter{table}{0}
\setcounter{equation}{0}

\newcommand{\hbAppendixPrefix}{A}
\renewcommand{\thesection}{S\arabic{section}}
\renewcommand{\thetable}{S\arabic{table}}
\renewcommand{\thefigure}{S\arabic{figure}}

\twocolumn[\centerline{\Large{\textbf{Supplemental Material}}}]

\vspace*{0.7cm}

This supplemental material introduces the calibration procedure used to estimate estimate the point spread functions (PSFs) (Sec.~\ref{sec:dpPSF}) along with the parameter range searching for our parametric dual-pixel (DP) PSF model (Sec.~\ref{sec:parametricPSF}). Next, another calibration procedure is described for estimating the radial distortion coefficients (Sec.~\ref{sec:radialDis}). Afterward, an ablation study is provided to demonstrate the effectiveness of each component added to our proposed recurrent dual-pixel deblurring architecture  (RDPD). Then, Sec.~\ref{sec:defocusMotion} provides a discussion about the difference between the defocus and motion blur. Additional qualitative results for different deblurring methods are also presented in Sec.~\ref{sec:qualRes}.

For further visual assessment, we provide animated examples that alternate between the left and right DP views of real and synthetic data in the ``animated-dp-views'' directory. We also provide animated examples of our deblurring results in the ``animated-results'' directory. Both ``animated-dp-views'' and ``animated-results'' are  located in the project GitHub repository: \url{https://github.com/Abdullah-Abuolaim/recurrent-defocus-deblurring-synth-dual-pixel}.

\section{PSFs on a real dual-pixel sensor}\label{sec:dpPSF}

\begin{figure}[t]
\vspace*{1.2cm}
\includegraphics[width=\linewidth]{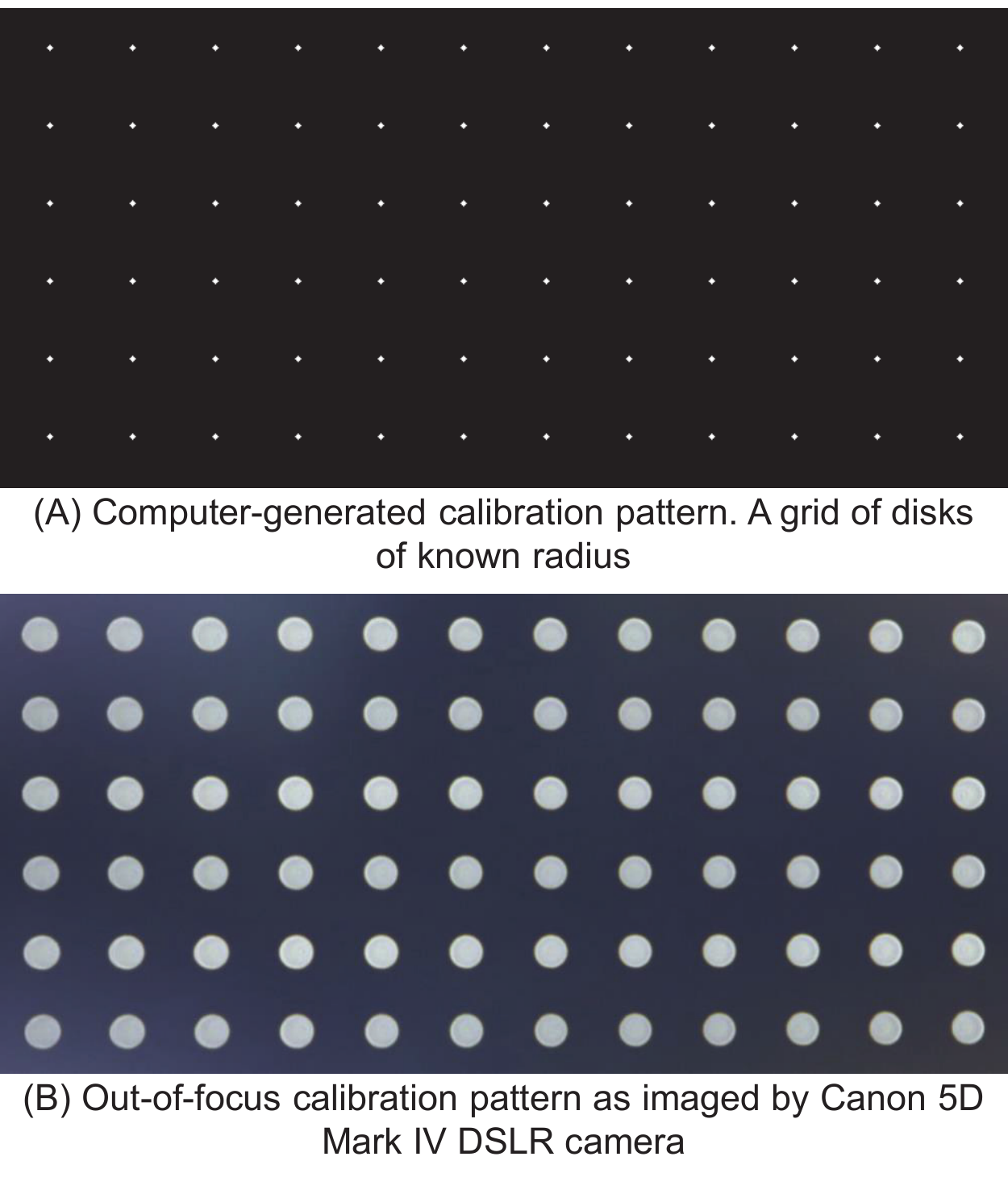}
\caption{Calibration patterns are used for estimating dual-pixel (DP) point spread functions (PSFs). A: our synthesized computer-generated pattern of a grid of small equal-size disks. B: the same calibration pattern as imaged by Canon 5D Mark IV DSLR camera. Note that the pattern captured in (B) is out-of-focus.}\label{fig:calibration}
\end{figure} 

As mentioned in Sec.~2.2 of the main paper, we aim to estimate more accurate and realistic DP-PSFs; thus, we follow the same calibration practice as in~\cite{joshi2008psf,mannan2016blur,punnappurath2020modeling}. The calibration pattern contains a grid of small disks with known radius and spacing as shown in Fig.~\ref{fig:calibration}-A. The pattern in Fig.~\ref{fig:calibration}-A is computer-generated, and we display it on a 27-inch LED display of resolution $1920\times1080$. Then, we use the Canon 5D Mark IV DSLR camera for our capturing procedure, as it facilitates reading out DP data. The LED display is placed parallel to the image plane and at a fixed distance of about one meter.

We captured many images by varying the camera parameters -- namely, focus distance, aperture size, and focal length -- covering a wide range of shape varying PSFs (an example image is shown in Fig.~\ref{fig:calibration}-B). Since we apply radial distortion to our synthetically generated data, we seek to estimate the least radially-distorted PSF, that in practice, is found close to the image center. Patches containing the disks are identified, and the center of the disks is estimated by finding these patches' centroid. The radius of computer-generated disks is a known fraction of the distance between disk centers.

\begin{figure*}[t]
\centering
\includegraphics[width=0.94\linewidth]{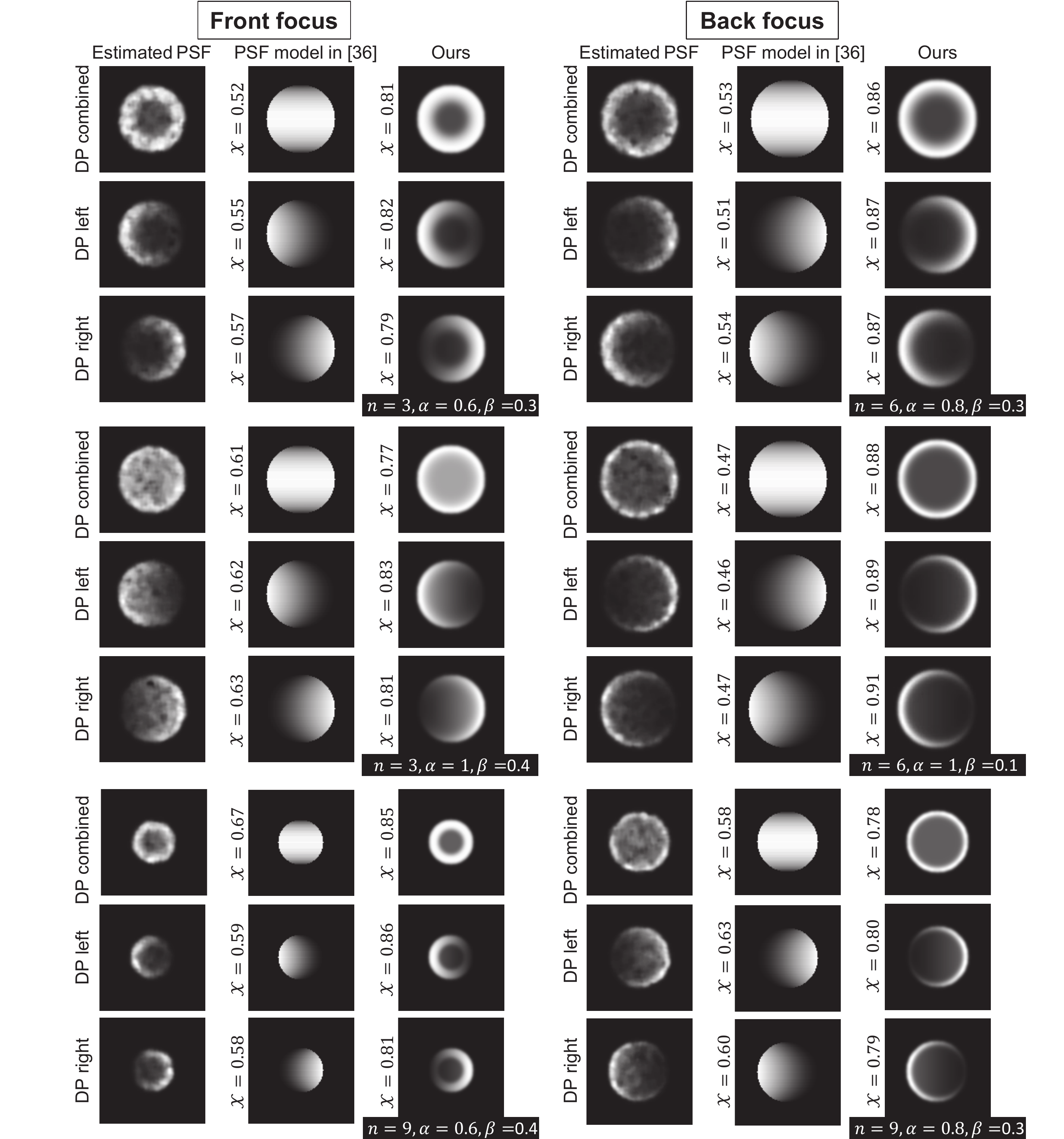}
\caption{The estimated point spread functions (PSFs) in comparison with the dual-pixel (DP) PSF model in~\cite{punnappurath2020modeling} and our parametric DP-PSF model. The PSFs for the DP combined, left, and right are estimated independently. The similarity with the estimated PSF is measured using the 2D cross-correlation $\mathcal{X}$(.) and is shown on the left for each case. The parameters (i.e., $n,\alpha,\beta$) used to generate our DP-PSF are shown below each case. Our parametric DP-PSF model obtains higher similarity with the estimated ones.}\label{fig:psfs}
\end{figure*} 

Similar to~\cite{mannan2016blur,punnappurath2020modeling}, we adopt a non-blind PSF estimation, in which the latent sharp disk $\mathbf{S}$ is known. Then, the PSF from the camera $\mathbf{E}$ is estimated using $\mathbf{S}$ and the corresponding blurred patch $\mathbf{B}$ by solving:
\begin{eqnarray}
&\argmin\limits_{\mathbf{E}} \sum_{i} \Big\| \mathbf{D}_i (\mathbf{S} \ast \mathbf{E} - \mathbf{B}) \Big\|_2^2+ \Big\|\mathbf{E} \Big\|_1, \nonumber \\
&\mbox{subject to} \ \mathbf{E} \geq \mathbf{0}, 
\end{eqnarray}
where $\mathbf{D}_i \in \{\mathbf{D}_x, \mathbf{D}_y, \mathbf{D}_{xx}, \mathbf{D}_{yy}, \mathbf{D}_{xy} \}$ denote the spatial vertical and horizontal derivatives. The $\ell_1$-norm of $\mathbf{E}$ encourages sparsity of the PSF entries. Additionally, another non-negativity constraint is imposed on the entries of $\mathbf{E}$. A wide range of PSFs for each DP view is estimated independently. Fig.~\ref{fig:psfs} shows examples of the estimated DP- PSFs.

\section{Parameter search for our dual-pixel PSFs}\label{sec:parametricPSF}
In Sec. 2.2 and 5 of the main paper, we also mentioned a mechanism to select the effective range of parameters (i.e.,  $n, \alpha, \beta, \kappa$) for our parametric DP-PSF modeling. Recall that $n$ is the Butterworth filter order, and $\alpha$ is used to control its 3db cutoff position.  $\beta$ is the filter lower bound scale, and $\kappa$ is the Gaussian smoothing factor.

Our goal is not to exactly match the measured PSFs found in cameras but rather to find representative PSFs that are very similar to what is estimated. The main reason is that we used a single camera, and as shown in Fig.~\ref{fig:psfs}, the estimated PSFs are noisy and not perfectly circular. This observation is expected due to camera-specific physical constraints like the positioning of the microlens, the depth of the sensor wells, and other optics manufacturing imperfections. Therefore, we limit the parameter search to a range of discrete values for each parameter sampled as:
\begin{eqnarray}
&n \in \{1,2,3,\dots,15\}, , \nonumber \\
&\alpha, \beta  \in \{0.1,0.2,\dots,1.0\}, \nonumber \\
&\kappa  \in \{0.14, 0.21,\dots,0.42\}.
\end{eqnarray}

Then, we perform a brute-force search in this bounded space by solving the following equation:
\begin{eqnarray}
&\argmin\limits_{n, \alpha, \beta, \kappa} \sum_{i} \Big\| \mathbf{D}_i (\mathbf{E} - \mathbf{H}(n, \alpha, \beta, \kappa))\Big\|_2^2, \end{eqnarray}
where $\mathbf{E}$ is the estimated PSF from a real camera and $\mathbf{H}(n, \alpha, \beta, \kappa)$ is our parameterized PSF. We found the parameters achieve the highest similarity at:
\begin{eqnarray}
&n \in \{3,6,9\}, \nonumber \\
&\alpha \in \{0.4,0.6,0.8,1.0\}, \nonumber \\
&\beta  \in \{0.1,0.2,0.3,0.4\}, \nonumber \\
&\kappa=0.14.
\end{eqnarray}

Given the best values we defined for each parameter, we can generate $48$ combinations of possible PSFs, and those represent our bank of DP-PSFs used to generate our synthetic DP data. Fig.~\ref{fig:psfs} shows examples of our parameterized PSFs in comparison with the estimated ones. Our PSFs demonstrate a much higher correlation with the estimated real PSFs compared to the DP-PSF model in~\cite{punnappurath2020modeling}. The similarity is measured using the 2D cross-correlation $\mathcal{X}(.)$.

Such comprehensive PSF calibration (Sec.~\ref{sec:dpPSF}) and parameter search (Sec.~\ref{sec:parametricPSF}) is not possible for smartphone cameras due to uncontrollable camera factors like aperture size and focal length. Additionally, up to our knowledge, only Pixel 3 and 4 smartphones allow direct access to the DP data, and the Pixel-DP API provided in~\cite{garg2019learning} does not facilitate manual focus, which limits us from controlling the focus distance as well. The work in~\cite{punnappurath2020modeling} estimated two PSFs from a Pixel 3 smartphone, and we found that they achieve a high correlation with our parametric PSF model.

\section{Radial distortion coefficients}\label{sec:radialDis}

\begin{figure}[t]
\includegraphics[width=\linewidth]{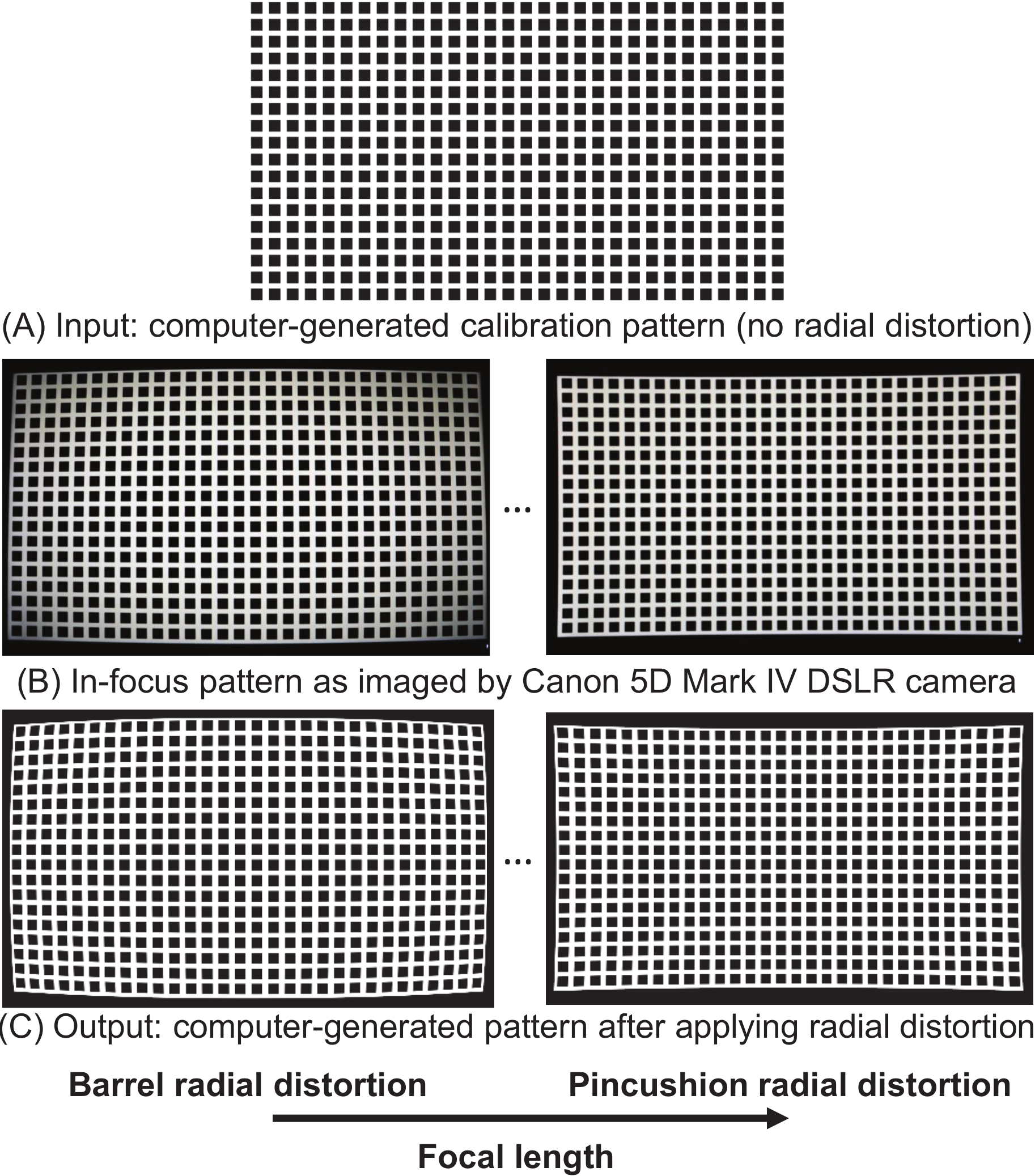}
\caption{Calibration pattern used for estimating the radial distortion coefficients. A: the input computer-generated pattern with no radial distortion applied. B: the same pattern as imaged by the Canon 5D Mark IV camera at different focal lengths. C: the computer-generated pattern after applying radial distortion.}\label{fig:radialDis}
\end{figure} 

Based on Sec. 2.3 of the main paper, radial distortion is applied for more realistic imagery since the input images and our parametric DP-PSFs are not radially distorted. To this aim, we use the division model~\cite{fitzgibbon2001simultaneous}, as follows:
\begin{eqnarray}\label{eq:radialDistortion}
(x_d,y_d)=(x_o,y_o) + \frac{(x_u-x_o,y_u-y_o)}{1+c_1 R^2+c_2 R^4+\cdots},
\end{eqnarray}
where $(x_u,y_u)$ and $(x_d,y_d)$ are the undistorted and distorted pixel coordinates respectively, and $c_i$ is the $i^\mathrm{th}$ radial distortion coefficient. $R$ is the radial distance from the image plane center $(x_o,y_o)$. This section introduces the calibration used to capture real-world radial distortion cases and is followed by the coefficient search procedure used to mimic real-world radial distortion. Recall that radial distortion is associated with zoom lenses and depends mainly on the camera's focal length. 

We synthesize a uniform pattern of squares, as shown in Fig.~\ref{fig:radialDis}-A. We follow the same setup described in Sec.~\ref{sec:dpPSF}. Still, with the following changes: (1) we capture an in-focus calibration pattern, (2) the focal length is changed across captures and the aperture remains fixed, (3) the distance between the display and camera is adjusted accordingly to make sure the full-resolution image is within camera's field of view, sine increasing the focal length introduces zoom/magnification effect, and (4) the focus distance is also adjusted accordingly to make sure the pattern is in focus. Fig.~\ref{fig:radialDis}-B shows examples of the calibration pattern as imaged by the Canon camera at different focal lengths.

We performed five captures at five different focal lengths ranging from min to max. Each is mapped to a focal length in our predefined parameter sets of the virtual five cameras -- namely: $\{4,5,6\},\{5,8,6\},\{7,5,8\},\{10,13,12\},\{22,10,30\}$ --- such that each set represents focal length, aperture size, and focus distance. With these five representative radial distortions that cover barrel as well as pincushion distortions, a brute-force search is performed to find the $c_i$ coefficients that satisfy the following:
\begin{eqnarray}\label{eq:radialDistortionSearch}
&\argmin\limits_{c_1,c_2,c_3} \mathcal{X}(\mathbf{I}_f,\mathbf{I}_d(c_1,c_2,c_3)),
\end{eqnarray}
where $\mathbf{I}_f$ is the calibration pattern as imaged by the Canon camera at a certain focal length $f$ (e.g., Fig.~\ref{fig:radialDis}-B). $\mathbf{I}_d(c_1,c_2,c_3)$ is the computer-generated input pattern but after applying radial distortion based on the coefficients $c_1,c_2,c_3$ (see example in Fig.~\ref{fig:radialDis}-C). Since we have few examples, $\mathbf{I}_f$ is re-centered manually to match $\mathbf{I}_d(c_1,c_2,c_3)$'s center. While Eq.~\ref{eq:radialDistortion} can be defined with more coefficients, we found three coefficients sufficient to approximate our real-world distortion examples. The final optimal five sets of coefficients are:
\begin{eqnarray}
&\{2\times10^{-2}, 2\times10^{-2}, 3\times10^{-2}\}, \nonumber \\
&\{8\times10^{-3}, 2\times10^{-3}, 2.2\times10^{-3}\}, \nonumber \\
&\{-4\times10^{-3}, 9\times10^{-4}, -9\times10^{-4}\}, \nonumber \\
&\{-7\times10^{-3}, -3.8\times10^{-3}, -3.6\times10^{-3}\}, \nonumber \\
&\{-8\times10^{-3}, -5\times10^{-3}, -4.5\times10^{-3}\}.
\end{eqnarray}

\section{Ablation study}\label{sec:ablationStudy}
This section investigates the usefulness of our synthetically generated DP data along with our novel recurrent dual-pixel deblurring architecture (RDPD) --- this is related to Sec. 4 and Sec. 5 of the main paper. To this aim, we divide the ablation study into two parts:
\begin{itemize}
  \item Explore the effectiveness of our DP data generator components (Sec.~\ref{sec:generatorComponents}), including: (1) our parametric DP-PSF model vs. the DP-PSF model presented in~\cite{punnappurath2020modeling}, (2) radially distorted DP data vs. undistorted ones, and (3) training with dual views vs. training with a single DP view.   
  \item Investigate the effectiveness of each component added to our RDPD model (Sec.~\ref{sec:RDPDComponents}), including: (1) the utility of adding the radial patch distance mask to the input and (2) our new edge loss vs. traditional Sobel loss.
\end{itemize}

Note that all subsequent experiments are conducted with variations of RDPD+. Each variation represents a single change, where the rest remains similar to RDPD+ as described in the main paper Sec. 5. All the variations are tested on Canon DP data from~\cite{abuolaim2020defocus}, since it is the only data that we can have access to real ground truth DP data and thus enables us to report quantitative results.

\subsection{Utility of DP data generator components}\label{sec:generatorComponents}
\begin{table}[t]
\centering
\caption{Results on indoor and outdoor scenes combined from the Canon DP dataset~\cite{abuolaim2020defocus}. Bold numbers are the best. RDPD+ (PSF~\cite{punnappurath2020modeling}) is a variation trained on DP data generated using the PSF model from~\cite{punnappurath2020modeling}. Our RDPD+, trained on DP data generated using our parametric DP-PSF, demonstrates +0.7db higher PSNR and reflects the power of our realistic PSF modeling.}
\scalebox{0.95}{
\begin{tabular}{l|c|c|c}
\toprule
\multicolumn{1}{c|}{\textbf{Variation}} & PSNR $\uparrow$ & SSIM $\uparrow$ & MAE $\downarrow$\\
\toprule
RDPD+ (PSF~\cite{punnappurath2020modeling}) & 24.69 & 0.752 & 0.044 \\ \hline

RDPD+ (our PSF) & {\bf 25.39} & {\bf 0.772}  & {\bf 0.040}\\ \hline
\end{tabular}
}
\label{tab:psfsAblation}
\end{table}

\noindent \textbf{Our parametric DP-PSF.}~While our parametric DP-PSF model already has a higher correlation with the estimated PSFs from real cameras, we further investigate the effect on RDPD+ when trained with data generated using other DP-PSFs. In this study, we compare our RDPD+ that is trained with DP data generated using our parametric DP-PSF model against the DP data generated using the DP-PSF model in~\cite{punnappurath2020modeling}.

The quantitative results reported in Table~\ref{tab:psfsAblation} demonstrates the power of training RDPD+ with DP data that is generated using our realistically modeled PSFs, where there is an increase in PSNR of +0.7db.

\begin{table}[t]
\centering
\caption{Results on indoor and outdoor scenes combined from the Canon DP dataset~\cite{abuolaim2020defocus}. Bold numbers are the best. RDPD+ trained with radially distorted DP data achieves +0.2db higher PSNR when tested on real images, in which applying radial distortion on the synthetically generated DP data helped RDPD+ to learn the spatially varying PSFs shapes found in real cameras.}
\scalebox{0.95}{
\begin{tabular}{l|c|c|c}
\toprule
\multicolumn{1}{c|}{\textbf{Variation}} & PSNR $\uparrow$ & SSIM $\uparrow$ & MAE $\downarrow$\\
\toprule
RDPD+ (w/o distortion) & 25.19 & 0.758 & 0.041 \\ \hline

RDPD+ (w/ distortion) & {\bf 25.39} & {\bf 0.772}  & {\bf 0.040}\\ \hline
\end{tabular}
}
\label{tab:radialDis}
\end{table}

\noindent \textbf{Radial distortion.}~For more realistic modeling, we considered applying radial distortion in the proposed DP data generator. In this study, we examine the proposed deblurring RDPD+ model's behavior when it is also trained with data that is not radially distorted. In Table~\ref{tab:radialDis}, we present the quantitative results of training RDPD+ with data generated with and without radial distortion. The results demonstrate the effectiveness of modeling the radial distortion during synthesizing the DP images, where RDPD+ trained with radially distorted data leads to a +0.2db PSNR gain.

\begin{table}[t]
\centering
\caption{Results on indoor and outdoor scenes combined from the Canon DP dataset~\cite{abuolaim2020defocus}. Bold numbers are the best. RSPD+: recurrent single-pixel deblurring trained with a single DP view (i.e., left view). RDPD+: trained with left and right DP views. Utilizing DP views to train our RDPD+ leads to +1.15db PSNR gain and is essential for better defocus deblurring.}
\scalebox{0.95}{
\begin{tabular}{l|c|c|c}
\toprule
\multicolumn{1}{c|}{\textbf{Variation}} & PSNR $\uparrow$ & SSIM $\uparrow$ & MAE $\downarrow$\\
\toprule
RSPD+ & 24.24 & 0.726 & 0.045 \\ \hline

RDPD+ & {\bf 25.39} & {\bf 0.772}  & {\bf 0.040}\\ \hline
\end{tabular}
}
\label{tab:singleDPview}
\end{table}

\noindent \textbf{Dual views vs. single view.}~ Following~\cite{abuolaim2020defocus}, we explore the effect of training RDPD+ with DP views vs. training with a single view (i.e., the traditional approach of single image deblurring~\cite{karaali2017edge,lee2019deep,shi2015just}). To this aim, we introduce a recurrent single-pixel deblurring model variant (RSPD+) and compare it with our proposed RDPD+ model. Quantitative results in Table~\ref{tab:singleDPview} demonstrates that training with DP views is crucial in order to perform better defocus deblurring where there is a +1.15db PSNR gain.

\subsection{Effectiveness of RDPD components}\label{sec:RDPDComponents}
\begin{table}[t]
\centering
\caption{Results on indoor and outdoor scenes combined from the Canon DP dataset~\cite{abuolaim2020defocus}. Bold numbers are the best. RDPD+ has an improved results when it is trained with the radial distance patch.}
\scalebox{0.9}{
\begin{tabular}{l|c|c|c}
\toprule
\multicolumn{1}{c|}{\textbf{Variation}} & PSNR $\uparrow$ & SSIM $\uparrow$ & MAE $\downarrow$\\
\toprule
RDPD+(w/o radial distance) & 25.00 & 0.756 & 0.042 \\ \hline

RDPD+(w/ radial distance) & {\bf 25.39} & {\bf 0.772}  & {\bf 0.040}\\ \hline
\end{tabular}
}
\label{tab:radialPatch}
\end{table}

\begin{figure}[t]
\centering
\includegraphics[width=\linewidth]{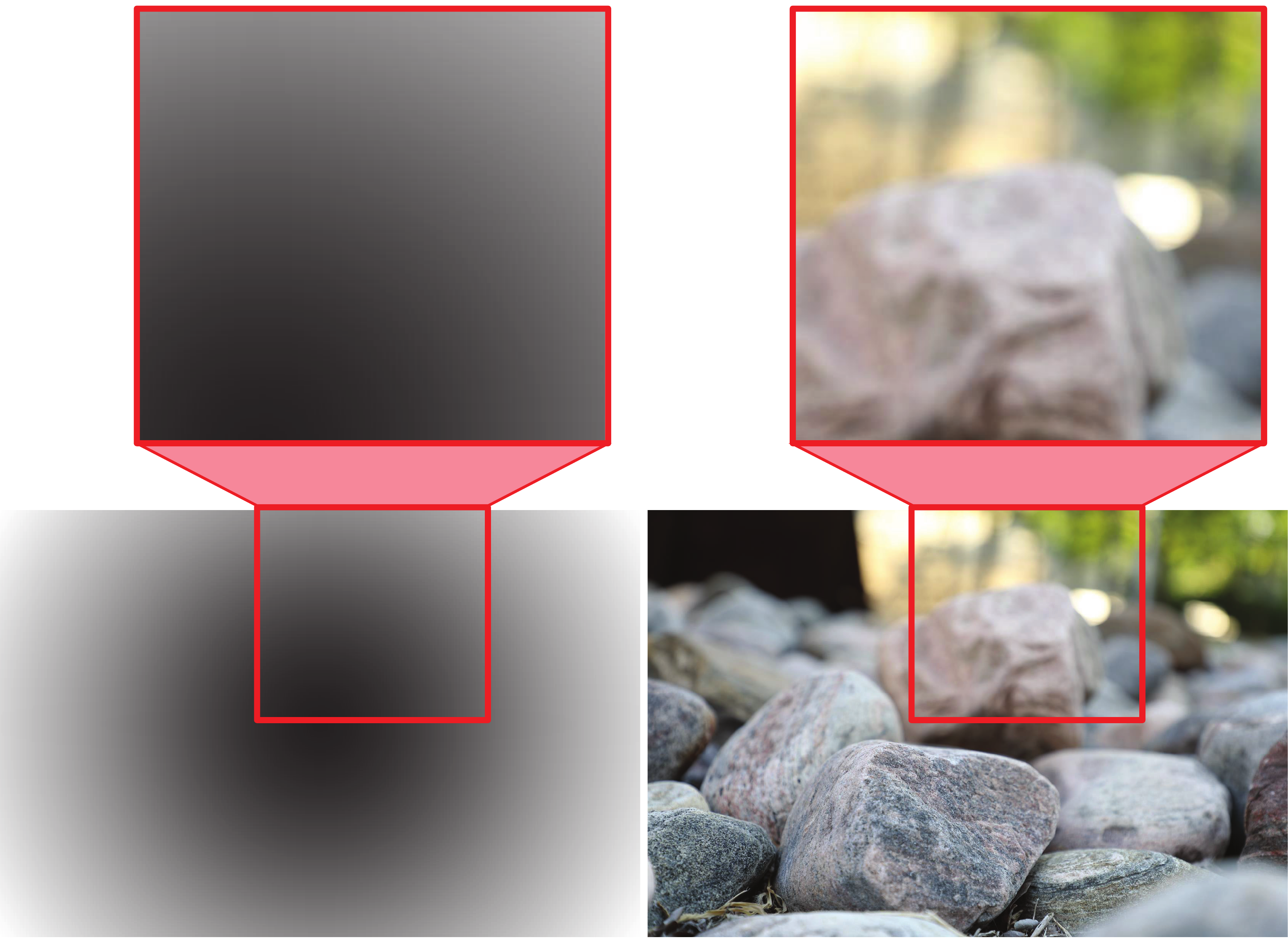}
\caption{The radial distance patch used to assist RDPD+ training and address the issue of patch-wise training.}\label{fig:radialDistance}
\vspace{-3mm}
\end{figure} 

\noindent \textbf{Radial distance patch for patch-wise training.}~We introduced training with the radial distance patch to feed each pixel's spatial location in the cropped patch and address the patch-wise training issue (i.e., the network does not see the full image or the relative position of the patch in the full image). Training in this manner is important as the PSFs are spatially varying in the radial direction away from the image center. Fig.~\ref{fig:radialDistance} shows an example of the cropped radial distance patch used to assist our training. To investigate the effectiveness of training with the radial distance patch, we also train RDPD+ without it and report the results in Table~\ref{tab:radialPatch}. RDPD+ has a gain in PSNR of +0.4db when trained with the radial distance patch.

\begin{table}[t]
\centering
\caption{Results on indoor and outdoor scenes combined from the Canon DP dataset~\cite{abuolaim2020defocus}. Bold numbers are the best. { \bf RDPD+(0)}: trained without the edge loss. { \bf RDPD+(1)}: trained with a $3\times3$ single-scale Sobel loss similar to~\cite{lu2019single}. { \bf RDPD+(3)}: trained with our three-scale edge loss. RDPD+ trained with our multi-scale edge loss has the best results for all metrics.}
\scalebox{0.95}{
\begin{tabular}{l|c|c|c}
\toprule
\multicolumn{1}{c|}{\textbf{Variation}} & PSNR $\uparrow$ & SSIM $\uparrow$ & MAE $\downarrow$\\
\toprule
RSPD+(0) & 25.06 & 0.765 & 0.042 \\ \hline

RSPD+(1)~\cite{lu2019single} & 25.11 & 0.763 & 0.042 \\ \hline

RDPD+(3) & {\bf 25.39} & {\bf 0.772}  & {\bf 0.040}\\ \hline
\end{tabular}
}
\label{tab:edgeLoss}
\end{table}
\noindent \textbf{Our multi-scale edge loss.}~As mentioned in Sec. 4 of the main paper, we introduced the multi-scale edge loss based on the Sobel gradient operator to recover sharper details at different edge sizes. To examine our edge loss function's effectiveness, we train RDPD+ with different variations of edge loss scales denoted as RDPD+($m$), where $m$ is the number of scales used. In particular, we introduce RDPD+(0) (trained without the edge loss), RDPD+(1) (trained with a $3\times3$ single-scale Sobel loss similar to~\cite{lu2019single}), and RDPD+(3) (trained with our three-scale edge loss). Table~\ref{tab:edgeLoss} shows the quantitative results, in which training with our multi-scale edge loss achieves the best results for all metrics.

\begin{table}[t]
\centering
\caption{Results on indoor and outdoor scenes combined from the Canon DP dataset~\cite{abuolaim2020defocus}. Bold numbers are the best. RDPD+ has about -0.5db PSNR drop in performance, when both our radial distance patch and multi-scale edge loss are removed from the RDPD+'s training.}
\scalebox{0.9}{
\begin{tabular}{l|c|c|c}
\toprule
\multicolumn{1}{c|}{\textbf{Variation}} & PSNR $\uparrow$ & SSIM $\uparrow$ & MAE $\downarrow$\\
\toprule
RDPD+(w/o both) & 24.90 & 0.754 & 0.042\\ \hline

RDPD+(w/ both) & {\bf 25.39} & {\bf 0.772}  & {\bf 0.040}\\ \hline
\end{tabular}
}
\label{tab:radialPatchEdgeLoss}
\vspace{-3mm}
\end{table}
\noindent \textbf{Effect of radial distance and edge loss together.}~We also examine the performance when our radial distance patch and  multi-scale edge loss are removed from the RDPD+'s training. Table~\ref{tab:radialPatchEdgeLoss} shows the results, where there is a PSNR drop of -0.5db, indicating the usefulness of the additional proposed components.

\section{Defocus vs. motion deblurring}\label{sec:defocusMotion}
While defocus and motion both lead to image blur, the physical formation and appearance of these two blur types are significantly different. Therefore, methods that solve for motion blur are not expected to perform well when applied to defocus deblurring. This is shown by Abuolaim et al.~\cite{abuolaim2020defocus}, where the motion deblurring method of Tao et al.~\cite{tao2018scale} that is evaluated on the same Canon test set achieves an average PSNR of $20.12$dB, which is significantly lower than ours (i.e., 25.39) and all other defocus deblurring methods.
\section{Additional qualitative results}\label{sec:qualRes}
\begin{figure*}[t]
\centering
\includegraphics[width=\linewidth]{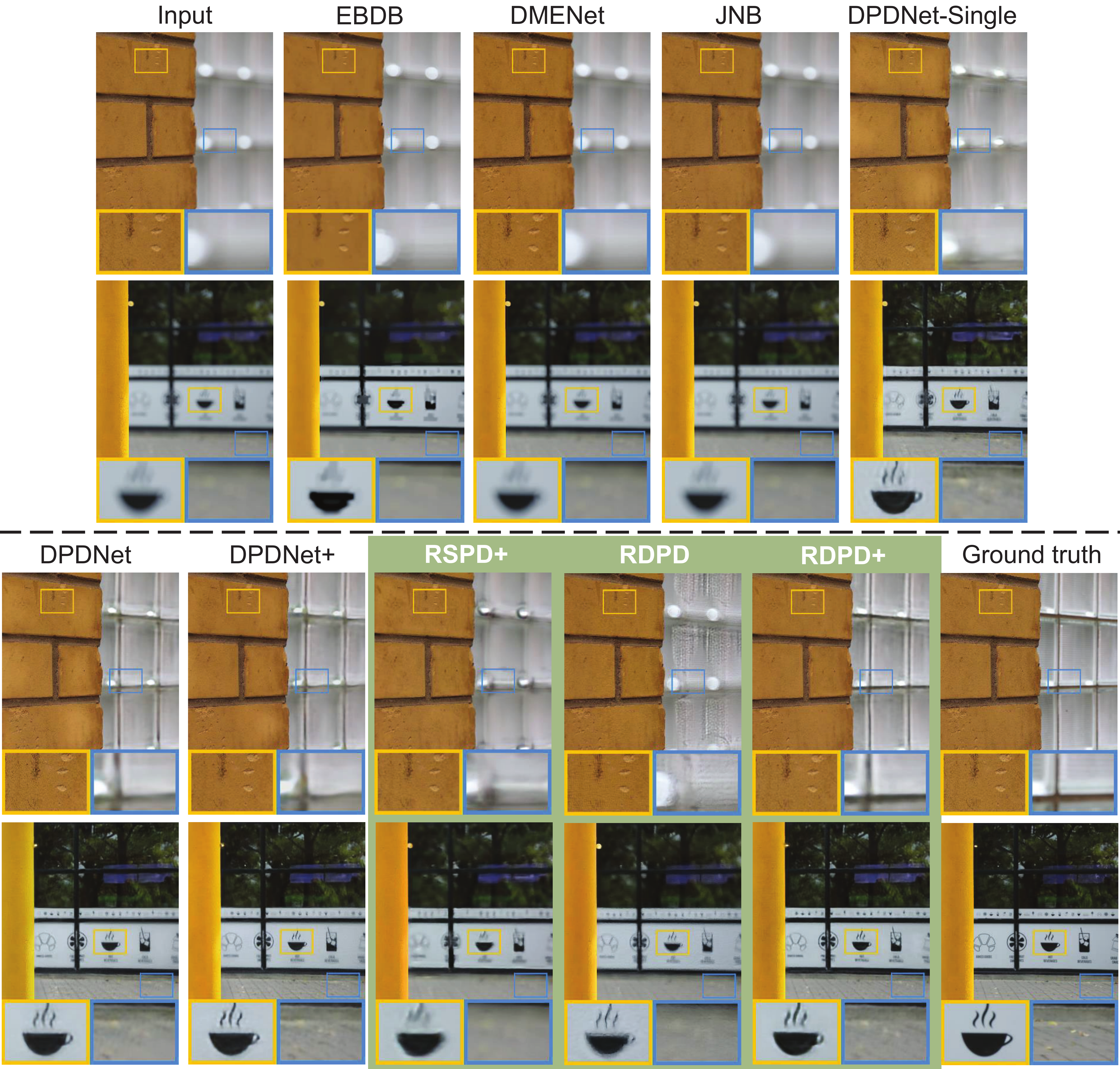}
\caption{Qualitative results on images from Canon dataset~\cite{abuolaim2020defocus}. We compare with other defocus deblurring methods: EBDB~\cite{karaali2017edge}, DMENet~\cite{lee2019deep}, JNB~\cite{shi2015just}, and DPDNet~\cite{abuolaim2020defocus}. RDPD is our method trained on synthetically generated DP data only. DPDNet+ and RDPD+ are trained on both Canon and synthetic DP data. DPDNet-Single and RSPD+ are trained on a single DP view (i.e., $\mathbf{I}_l$). In general, RDPD and RDPD+ are able to recover more
image details. Interestingly, RDPD trained on synthetic data generalizes well to real data from Canon camera.}\label{fig:qualCanon1}
\end{figure*}

\begin{figure*}[t]
\centering
\includegraphics[width=\linewidth]{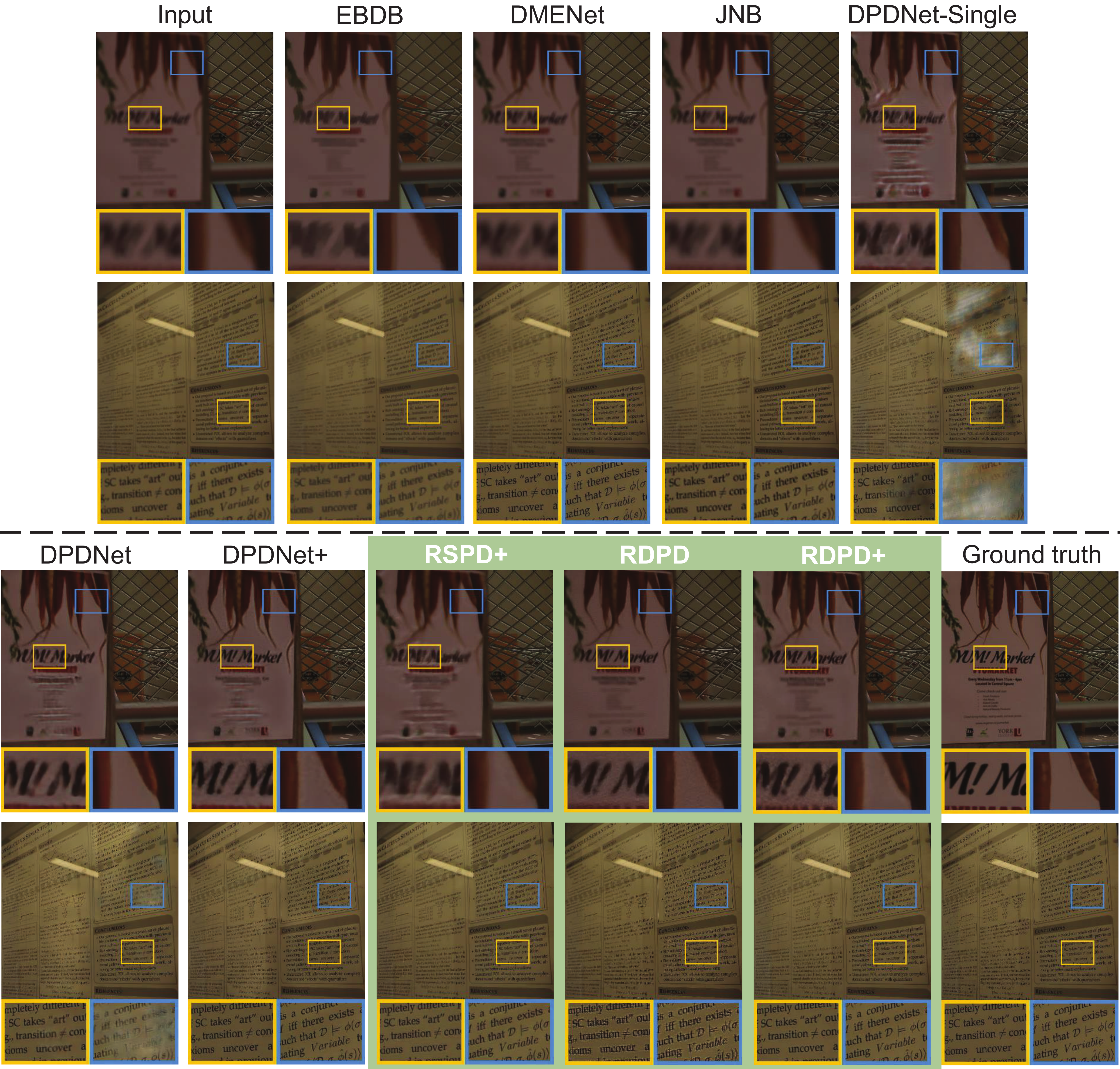}
\caption{Qualitative results on images from Canon dataset~\cite{abuolaim2020defocus}. We compare with other defocus deblurring methods: EBDB~\cite{karaali2017edge}, DMENet~\cite{lee2019deep}, JNB~\cite{shi2015just}, and DPDNet~\cite{abuolaim2020defocus}. RDPD is our method trained on synthetically generated DP data only. DPDNet+ and RDPD+ are trained on both Canon and synthetic DP data. DPDNet-Single and RSPD+ are trained on a single DP view (i.e., $\mathbf{I}_l$). In general, RDPD and RDPD+ are able to recover more
image details. Interestingly, RDPD trained on synthetic data generalizes well to real data from Canon camera.}\label{fig:qualCanon2}
\end{figure*}

\begin{figure*}[t]
\centering
\includegraphics[width=\linewidth]{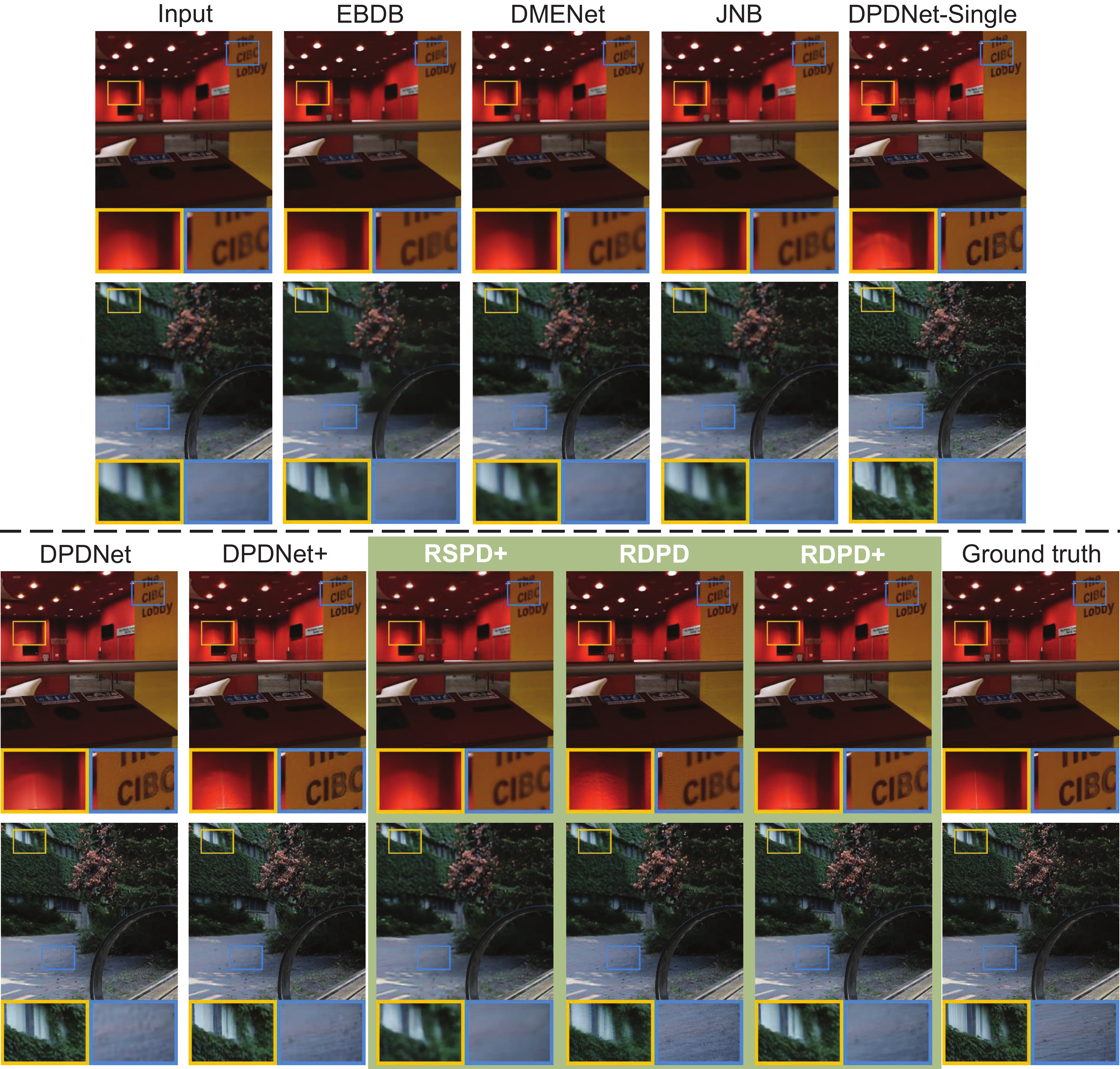}
\caption{Qualitative results on images from Canon dataset~\cite{abuolaim2020defocus}. We compare with other defocus deblurring methods: EBDB~\cite{karaali2017edge}, DMENet~\cite{lee2019deep}, JNB~\cite{shi2015just}, and DPDNet~\cite{abuolaim2020defocus}. RDPD is our method trained on synthetically generated DP data only. DPDNet+ and RDPD+ are trained on both Canon and synthetic DP data. DPDNet-Single and RSPD+ are trained on a single DP view (i.e., $\mathbf{I}_l$). In general, RDPD and RDPD+ are able to recover more image details. Interestingly, RDPD trained on synthetic data generalizes well to real data from Canon camera.}\label{fig:qualCanon3}
\end{figure*}

\begin{figure*}[t]
\centering
\includegraphics[width=\linewidth]{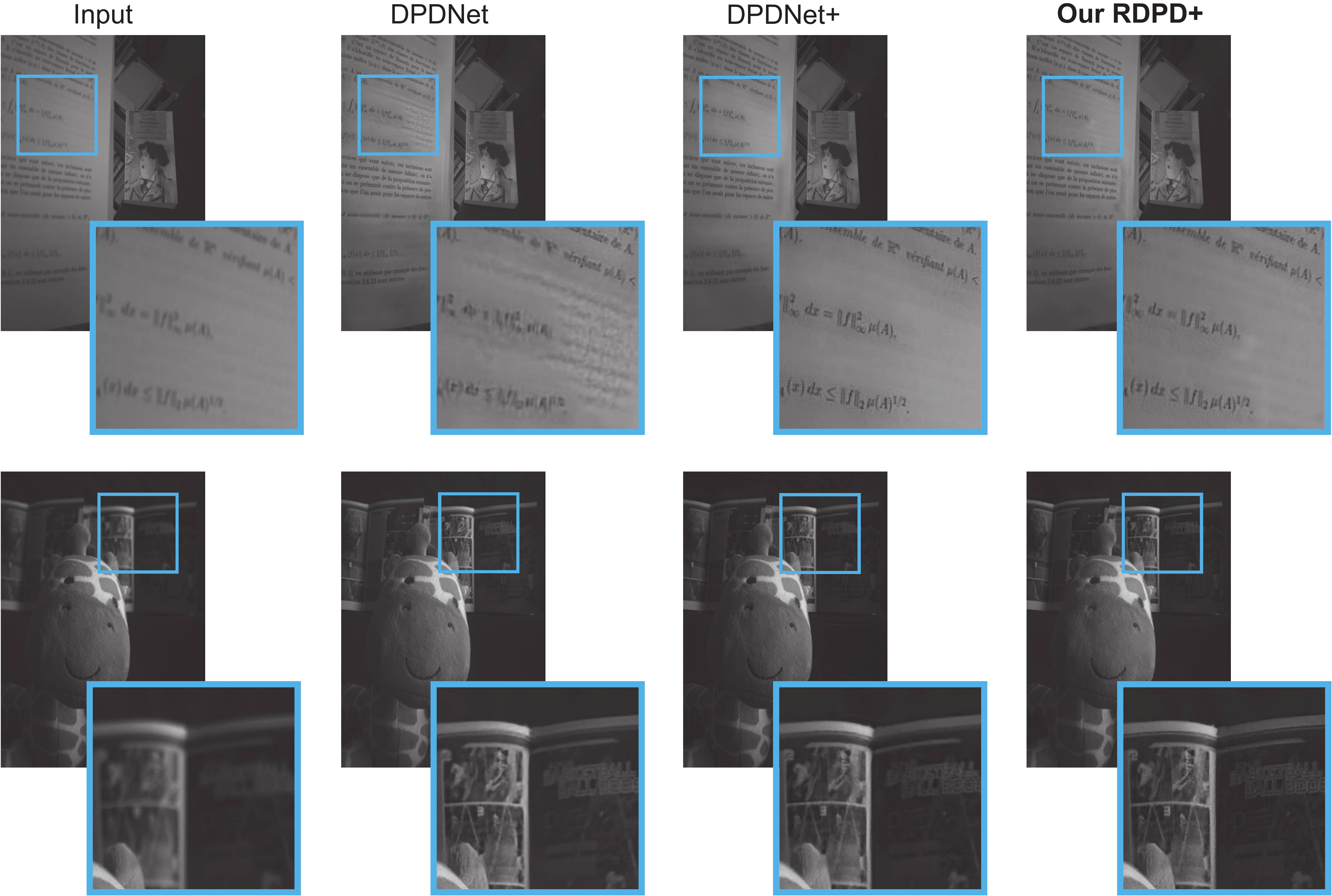}
\caption{Qualitative results on images captured by Pixel smartphone. DPDNet~\cite{abuolaim2020defocus} is trained on Canon DP data only. DPDNet+ and RDPD+ are trained on both Canon and synthetic DP data. In general, RDPD+ is able to recover more image details. Interestingly, DPDNet+ achieves better results when it is trained with our synthetic data augmented. Note that there is no ground truth sharp image for Pixel smartphone because smartphones have a fixed aperture.  As a result,  a narrow-aperture image cannot be captured to serve as a ground truth image. Additionally, we note that the DP data currently available from the Pixel smartphones are not full-frame but are limited to only one of the green channels in the raw-Bayer frame.}\label{fig:qualpixel}
\end{figure*}

As mentioned in Sec. 5 of the main paper, we provide more qualitative results for all existing defocus deblurring methods including a variations of ours. The methods are: the DP deblurring network (DPDNet)~\cite{abuolaim2020defocus}, the edge-based defocus blur (EBDB)~\cite{karaali2017edge}, the defocus map estimation network (DMENet)~\cite{lee2019deep}, and the just noticeable blur (JNB)~\cite{shi2015just} estimation. Fig.~\ref{fig:qualCanon1}, Fig.~\ref{fig:qualCanon2}, and Fig.~\ref{fig:qualCanon3} show more qualitative results on images from the Canon dataset~\cite{abuolaim2020defocus}. Fig.~\ref{fig:qualpixel} shows results on images from a Pixel smartphone.

\clearpage
\clearpage

{\small
\bibliographystyle{ieee_fullname}
\bibliography{arxiv_2021_synth_dp_defocus_deblurring}
}

\end{document}